\documentclass[%
 article,
nofootinbib,
amsmath,amssymb,
aps,
]{revtex4-2}

\usepackage{graphicx}
\usepackage{dcolumn}
\usepackage{bm}

\usepackage[caption=false]{subfig}
\usepackage{xcolor}
\usepackage{newtxtext,newtxmath}
\usepackage[T1]{fontenc}
\usepackage{tabularx}
\usepackage{cancel}
\usepackage{hyperref}
\usepackage{cleveref}
\usepackage{multirow}
\usepackage{float}
\usepackage{colortbl}

\crefformat{figure}{Fig.~#2#1#3}
\crefformat{equation}{equation~#2#1#3}
\crefformat{section}{section~#2#1#3}
\crefformat{table}{Tab.~#2#1#3}
\crefformat{appendix}{appendix~#2#1#3}
\crefmultiformat{figure}{Figs.~#2#1#3}{~and~#2#1#3}%
    {,~#2#1#3}{,~#2#1#3}
\crefrangeformat{figure}{Figs.~(#3#1#4--#5#2#6)}

\definecolor{LightCyan}{rgb}{0.88,1,1}

\begin{document}

\title{Astrophysical Constraints from the SARAS3 \\ non-detection of the Cosmic Dawn Sky-Averaged 21-cm Signal}

\author{H. T. J. Bevins$^{1,*}$, 
A. Fialkov$^{2,3}$, 
E. de Lera Acedo$^{1, 2}$, 
W. J. Handley$^{1, 2}$, 
S. Singh$^{4}$, 
R. Subrahmanyan$^{5}$, 
and R. Barkana$^{6, 7, 8}$
\\
$^{1}$Astrophysics Group, Cavendish Laboratory, J. J. Thomson Avenue, Cambridge, CB3 0HE, UK\\
$^{2}$Kavli Institute for Cosmology, Madingley Road, Cambridge CB3 0HA, UK \\
$^{3}$Institute of Astronomy, University of Cambridge, Madingley Road, Cambridge CB3 0HA, UK \\
$^{4}$ Raman Research Institute, C V Raman Avenue, Sadashivanagar, Bangalore 560080, India \\
$^{5}$ Space \& Astronomy CSIRO, 26 Dick Perry Ave, Kensington WA 6151, Australia \\
$^{6}$ School of Physics and Astronomy, Tel-Aviv University, Tel-Aviv, 69978, Israel \\
$^{7}$ Institute for Advanced Study, 1 Einstein Drive, Princeton, New Jersey 08540, USA \\
$^{8}$ Department of Astronomy and Astrophysics, University of California, Santa Cruz, CA 95064, USA \\
$^{*}$ htjb2@cam.ac.uk
}

\date{\today}

\begin{abstract} 
\textbf{Observations of the redshifted 21-cm line of atomic hydrogen have resulted in several upper limits on the 21-cm power spectrum and a tentative detection of the sky-averaged signal at $z\sim17$. Made with the EDGES Low-Band antenna, this claim was recently disputed by the SARAS3 experiment, which reported a non-detection and is the only available upper limit strong enough to constrain cosmic dawn astrophysics. We use these data to constrain a population of radio-luminous galaxies $\sim 200$ million years after the Big Bang ($z\approx 20$). We find, using Bayesian data analysis, that the data disfavours (at 68\% confidence) radio-luminous galaxies in dark matter halos with masses of $4.4\times10^{5}$~M$_\odot \lesssim M \lesssim 1.1\times10^{7}$~M$\odot$ (where $M_\odot$ is the mass of the Sun) at $z = 20$ and galaxies in which $>5\%$ of the gas is converted into stars. The data disfavour galaxies with radio luminosity per star formation rate of $L_\mathrm{r}/\mathrm{SFR} \gtrsim 1.549 \times 10^{25}$~W Hz$^{-1}$M$_\odot^{-1}$ yr at 150~MHz, a thousand times brighter than today, and, separately, a synchrotron radio background in excess of the CMB by $\gtrsim 6\%$ at 1.42~GHz.}
\end{abstract}
\maketitle

\section{Probing Cosmic Dawn}\label{sec1}

Understanding the early Universe, when the first stars and galaxies formed, is one of the major science goals of a number of new observatories. The recently launched JWST will directly image these early galaxies in deep near-infrared surveys. Its increased sensitivity, in comparison to the previous generation of telescopes, will allow JWST to target faint high-redshift galaxies existing during the first few hundred million years of cosmic history all the way out to the cosmological redshift of $z\sim 20$ \cite{Windhorst_JWST_2006}. Future confirmed and proposed X-ray missions, such as ATHENA~\cite{Athena}, LYNX~\cite{Lynx} and AXIS \cite{axis}, will supplement this exploration by observing the hot gas in the Universe.
Radio telescopes aim at complementing this picture by mapping the neutral intergalactic gas across the first billion years of cosmic history via observations of the 21-cm spin-flip transition of atomic hydrogen seen against the radio background radiation, which is usually assumed to be the Cosmic Microwave Background~(CMB) \cite{Madau, Mesinger_2019}. 

Upper limits on the 21-cm signal from the Epoch of Reionization~(EoR, $z\sim6-15$) are already available measured by both the radiometers \cite{ EDGES_high_band_experimental_paper_2017, SARAS2_radiometer_2018}, which probe the sky-averaged (global) 21-cm signal, and large interferometric arrays \cite{HERA_2017, LOFAR_current_EoR_2018, LEDA_2018, Trott_mwa_2020, Gehlot_lofar_2019} targeting fluctuations. These data have recently allowed constraints to be derived on the astrophysical processes at the EoR \cite{Singh_saras2_2017, Singh_saras2_2018, Monsalve_2019, Mondal_LOFAR_2020, Ghara_MWA_2021, LOFAR2021, HERA, SARAS2} and are in a broad agreement with other probes of reionization history such as high redshift quasars and galaxies \cite{Mesinger_2010, Schroder_2012, Ouchi_2018, Morales_2021, Greig_2022}.

The observational status of Cosmic Dawn~(CD) signal originating from higher redshifts ($z\sim15-30$) is more intriguing: the EDGES Low-Band collaboration reported a tentative detection of an absorption profile at $z\sim 17$ \citep{EDGES2018} which is at least two times deeper than what is predicted by conventional theoretical modelling \citep{Cohen_charting_2017, Reis_sta_2021}. Such a strong signal implies either an existence of an excess radio background above the CMB \citep{Feng2018,EDGES2018} or a non-standard thermal history of the intergalactic gas \cite{BarkanaDM2018,EDGES2018}.
The cosmological origin of this signal was recently disputed by the SARAS3 collaboration, who conducted an independent experiment and reported a non-detection of the EDGES best-fit profile in their data \cite{SARAS3_spectrometer_2020, SARAS3_antenna_2021, SARAS_reciever_2021, SARAS3}.
It has also been shown that the reported EDGES signal can partially be explained by invoking sinusoidal instrument systematics \cite{Hills2018, Singh2019, Bradley2019, Sims2020}, however, additional efforts are being made to verify the EDGES detection and to make independent measurements of the 21-cm signal from CD both with interferometers \cite{Mellema_SKA_2013, Zarka_nenuFar_2018, AARTFAAC_2020, HERA} and radiometers \cite{8879199, MIST, LEDA_2018}. 

The non-detection by SARAS3 of the EDGES profile increases the likelihood of the anomalous absorption feature being non-cosmological and brings the focus back to the more conventional astrophysical scenarios. 
In this work, we use the SARAS3 data to provide  constraints on the astrophysical processes at CD. We consider a potential population of high-redshift radio-luminous galaxies which contribute to the radio background radiation, thus affecting the 21-cm signal. In general, radio galaxies are expected in standard astrophysical scenarios \cite{Mirocha2019, Reis2020}, it is only for extremely high values of their radio luminosity that their contribution is large enough to explain the EDGES signal \cite{EDGES2018, Feng2018, Ewall2018, Jana2018, Mirocha2019, Fialkov2019, Reis2020}. Here, we consider a wide selection of models varying astrophysical properties of high-redshift galaxies over a broad range. We repeat our analysis for two additional scenarios (shown in \textit{Supplementary Material}): one with the CMB as the radio background radiation \cite{Reis_sta_2021} and the other with a phenomenological synchrotron radio background in addition to the CMB \cite{Fialkov2019}.  

In \cref{sec:data} we discuss in more detail the SARAS3 data. \Cref{sec:modelling} introduces the different modelled components in our analysis and discusses how we determine constraints on the astrophysical processes at CD, with further details given in \textit{Methods}. Our constraints on high-redshift radio galaxies are discussed in \cref{sec:results}. We conclude in \cref{sec:conclusions}. Additional astrophysical models are discussed in \textit{Supplementary Material}. 

\section{Data} \label{sec:data}

SARAS3 is a radiometer based on a monocone antenna that has made observations of the sky from a location in Southern India in the band $43.75-87.5$ MHz, targeting the cosmological 21-cm signal from $z\sim 15-32$ 
\cite{SARAS3_spectrometer_2020,SARAS3_antenna_2021,SARAS_reciever_2021}. The experiment is the first global 21-cm experiment of its kind to take observations whilst floating on a body of water which is expected to improve the total efficiency of the antenna. The total efficiency quantifies how the sky radiation is coupled to the antenna, including losses in the local environment of the antenna such as ground or water beneath the antenna. \cite{SARAS3_antenna_2021} and prevent the introduction of non-smooth systematics caused by stratified ground emission that can impede the detection of a global signal \cite{SARAS2}.

15 hours of observations were integrated in the frequency range $55-85$~MHz~($z\sim 15 - 25$), reduced after radio frequency interference filtering, with corrections made for emission from the water beneath the antenna and receiver noise temperature. The data were then appropriately scaled, given an estimate of the total efficiency,
to produce an average measurement of the sky temperature, $T_\mathrm{sky}$,  
which we expect to be the sum of the Galactic and extra-Galactic foregrounds, $T_\mathrm{fg}$, and the cosmological 21-cm signal, $T_{21}$.

Previously, a log-log polynomial foreground model was fitted to the data in combination with the phenomenological best-fit EDGES absorption profile multiplied by a scale factor, $s$, using a Markov Chain Monte Carlo analysis.
The data were shown to reject the presence of the EDGES signal with 95.3\% confidence and a series of EDGES-like signals, representing the likelihood distributions of uncertainties in the profile parameters, were rejected with a significance of 90.4\% \cite{SARAS3}. The SARAS3 measurement of the sky-averaged radio spectrum thus represents a non-detection of the EDGES absorption feature, with the potential to constrain astrophysical scenarios that result in signals larger in magnitude than the instrument noise.

\section{The Cosmological 21-cm Signal}
\label{sec:modelling}

To provide constraints on the astrophysical processes at CD using the SARAS3 data, we need to model the global 21-cm signal. Theoretical predictions of the signal are made difficult owing to the non-local impact of the non-uniform radiative fields produced by a distribution of luminous sources. Either numerical or semi-numerical methods are required to calculate the three-dimensional 21-cm signal and evolve it with time. The global signal can then be calculated as the spatial average, although one-dimensional radiative transfer codes are also used to calculate the global signal \cite{Mirocah:2020}. As astrophysical processes at high redshifts are poorly understood, a range of theoretical predictions for the 21-cm signal need to be computed for different astrophysical scenarios, which can then be constrained by data. In this work, we use a semi-numerical method \cite{Visbal2012, Fialkov2013, Fialkov2014, Fialkov2014Natur, Reis_sta_2021} to calculate the 21-cm signal from an evolving simulated Universe.

The calculation takes into account important processes that shape the 21-cm signal: The baryonic matter in the early Universe is predominantly composed of atomic hydrogen and, as the first stars and black holes form at $z\sim 30$ \cite{Fialkov2012, Klessen2019}, they affect both the total intensity and the fluctuations of the hydrogen signal. Radiation from the first stars in the Lyman band plays a fundamental role as it enables observations of the 21-cm signal against the radio background by coupling the characteristic temperature of the spin-flip transition, the spin temperature $T_\mathrm{S}$, to the kinetic temperature of the gas, $T_\mathrm{K}$ \cite{Wouthuysen, Field}. At CD, typically, radiation is warmer than the gas, and the signal appears as an absorption feature against the radio background. Heating of the intergalactic medium subsequently raises the gas temperature to and, perhaps, above the radio background \cite{Fialkov2014, Fialkov2014Natur, Reis_sta_2021}, resulting in emission at low redshifts. This evolution culminates at the EoR when ultraviolet radiation from stars ionizes the neutral hydrogen in the intergalactic medium and the signal disappears. If it exists at high redshifts, any additional radio background above the CMB would contribute to the 21-cm signal by increasing $T_\mathrm{rad}$ and, thus, deepening the absorption profile \cite{Feng2018, Ewall2018, Mirocha2019,Fialkov2019, Reis2020}.

For a specified set of astrophysical parameters, each simulation can take a few hours to produce the desired global 21-cm signal. However, in order to derive parameter constraints from real data, a multitude of such signals needs to be created probing the vast astrophysical parameter space. The application of machine learning to the problem of signal modelling is common in the analysis of data from 21-cm experiments \cite{Mondal_LOFAR_2020, HERA, Monsalve_2019, SARAS2} and allows different physical signal models to be generated quickly in computationally intensive fitting algorithms. Starting from a set of the simulated signals, we use neural networks \cite{Bevins_globalemu_2021} to interpolate the astrophysical parameter space (see \textit{Methods} and \cref{tab:networks} for a discussion of the network training and accuracy).

To extract constraints on the global signal, we also need to model the foreground in the data. We do this using the same log-log polynomial as in the previous analysis of the SARAS3 data  \cite{SARAS3} for consistency (see \textit{Methods}).

The theoretical CD 21-cm signal is sensitive to the process of star formation, thermal history and the temperature of the radio background radiation (see \textit{Methods}). The root mean squared~(RMS) of the appropriately weighted SARAS3 residuals after foreground modelling and removal is 213~mK at their native spectrum resolution of 61~kHz \cite{SARAS3}. Signals that are within the sensitivity of the instrument and would have been detected are those with deeper absorption troughs that have strong variation within the band. Typically, such signals are created in scenarios with high intensity of the Ly-$\alpha$ photons, corresponding to a combination of low minimum virial circular velocities $V_c$ and high star formation efficiencies $f_*$, and a strong contrast between the gas temperature and the temperature of the radio background, i.e. for low values of the X-ray production efficiency $f_X$ and high radio production efficiencies $f_\mathrm{radio}$. Therefore, we expect the SARAS3 data to constrain these model parameters. In our analysis, we marginalize parameters, including the CMB optical depth, $\tau$, and the mean free path of ionizing photons, $R_\mathrm{mfp}$, that determine the structure of the signal during the EoR because they are not relevant to the SARAS3 band.

We use the Bayesian nested sampling algorithm to perform our model fitting \cite{skilling_nested_2004} (see \textit{Methods}).

\section{Results} \label{sec:results}

\begin{figure}[h!]
    \centering
   \includegraphics{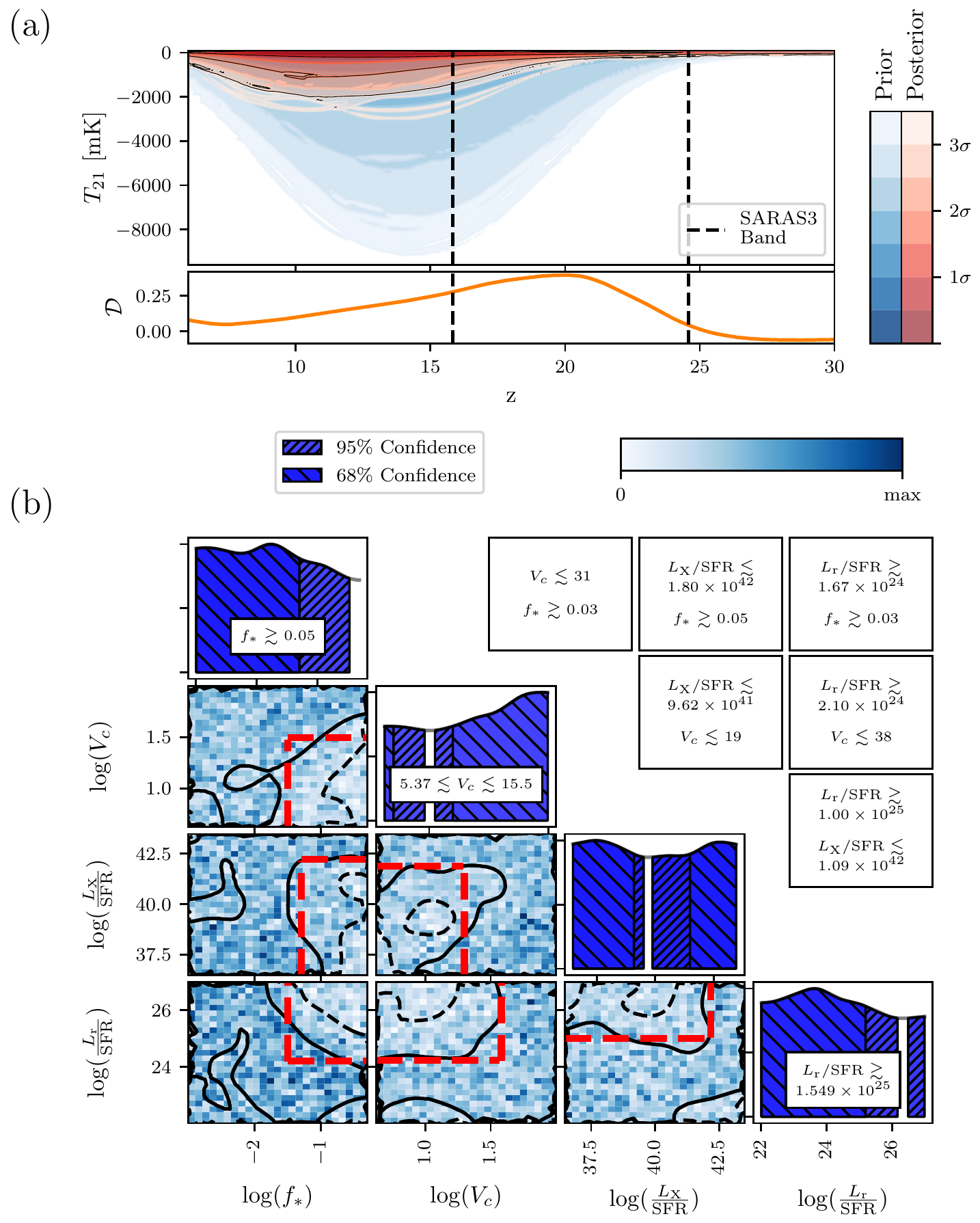}
   \caption{\textbf{SARAS3 constraints on high-redshift radio galaxies.}} 
    \label{fig:fradio_results}
\end{figure}

In this section, we discuss the SARAS3 constraints in the redshift range $z\sim 15-25$ on a population of high-redshift radio-luminous galaxies and show the main results in \cref{fig:fradio_results}. Constraints on models with the CMB-only background and an excess radio background from a phenomenological synchrotron source \cite{Fialkov2019} are discussed in  \textit{Supplementary Material}. 

We calculate the posterior distribution (using \cref{eq:posterior} in \textit{Methods}), which is a multivariate probability distribution for the thirteen parameters that describe the foreground~(seven polynomial coefficients), noise~(one parameter, the standard deviation of the Gaussian noise in the data, see \textit{Methods}) and the cosmological 21-cm signal~(five astrophysical parameters, $f_\mathrm{radio}$, $f_*$, $V_c$, $f_X$, and $\tau$ the CMB optical depth). We then marginalize over the foreground parameters, noise and $\tau$ which allows us to calculate the likelihood of different astrophysical 21-cm signals. The standard deviation of the noise is shown to be independent of the astrophysical parameters in \cref{fig:noise} for excess background models. We subsequently derive limits on the parameters related to star formation, heating, and the excess radio background above the CMB. In \textit{Methods} and \textit{Supplementary Material}, we discuss the foreground model in more detail \cite{SARAS3}. Although we note here that in fits with both a foreground and 21-cm signal model we found no correlation between the two sets of parameters as can be seen in \cref{fig:correlations}.

We find no evidence for an astrophysical signal in the data. The foreground-only fit consistently has a larger log-evidence by approximately 10 log units when compared to fits with signal profiles (i.e. $\mathcal{Z}_{M_1} > \mathcal{Z}_{M_2}$ as described in \textit{Methods}). Therefore,  any cosmological signal present in the data is likely to be undetectable in the residuals (with RMS of   $\approx 213$~mK), after foreground modelling and subtraction. Since the predicted amplitude of the 21-cm signal is expected to be lower than $\lesssim 165$ mK \cite{Reis_sta_2021} in the case of the standard scenario with the CMB as the only source of the background radiation, the SARAS3 constraints on this scenario are very weak (see \textit{Supplementary Material}). However, in the case of an excess radio background the predicted signals can be much stronger, allowing us to disfavour regions of the astrophysical parameter space. 

Panel (a) of \cref{fig:fradio_results} shows the contraction from the initial  set of possible 21-cm signals (prior, blue region, details of the astrophysical priors can be found in \textit{Methods} and \cref{tab:priors}) to the set of scenarios that are allowed by the data (i.e., functional posterior, shown in red). The functional posterior and prior is produced by taking the posterior samples returned from our nested sampling run and a representative set of prior samples are transformed into realizations of the global 21-cm signal using the trained neural networks. The figure illustrates that, as anticipated, the deepest signals and the signals with the strongest variation in the SARAS3 band, $z = 15-25$, are disfavoured. Signals with a modest variation within the band and signals with minima at $z \lesssim 15$ are indistinguishable from the foregrounds and, thus, cannot be ruled out. 

We can estimate the quantitative contribution of the SARAS3 measurement to our understanding of the high-redshift Universe by calculating the information gain using the Kullback-Leibler~(KL) divergence \cite{kullback_information_1951}, $\mathcal{D}$,  between the functional prior and posterior (bottom of panel (a)). The KL divergence by definition has arbitrary scaling but should always be $\gtrsim 0$. We see that $\mathcal{D}$ is highest, meaning the information gain is largest, at $z\sim 20$ which corresponds to the middle of the SARAS3 band. Owing to the dependence of the 21-cm signal on the star formation and heating histories, we see that the constraining power of the SARAS3 measurement extends to lower redshifts outside the SARAS3 frequency band (KL divergence is non-vanishing). On the other hand,  $\mathcal{D}$  is approximately zero at $z\gtrsim 30$ where the global 21-cm  signals are dominated by cosmology rather than astrophysics.

We next consider the astrophysical parameter constraints and show the corresponding 1D and 2D posteriors in panel (b) of \cref{fig:fradio_results}. The visualization of the constraints on the signal parameters is non-trivial and is achieved in 2D and 1D via marginalization. Marginalization involves integrating out of the posterior distribution the dependence on the $N-1$ or $N-2$ parameters to leave 1D and 2D distributions for the astrophysical parameters. Key numerical results and comparison with SARAS2 \cite{SARAS2} and HERA \cite{HERA} constraints are summarized in \cref{tab:numbers}. To guide the eye, we show the approximate 2D constraints (red dashed lines) roughly corresponding to the 68\% confidence contours (solid black lines) and list these limits in the figure (inverted triangle table). Note that there are regions of low probability outside these guides. From the figure it is clear that SARAS3 data most strongly constrain the process of primordial star formation (clear trends in the 1D posterior probabilities of $V_c$ and $f_*$) and the strength of the radio background (limits on the radio luminosity per star formation rate at 150~MHz, $L_\mathrm{r}/\mathrm{SFR}$) with somewhat weaker sensitivity to the heating process, which, however, is clearly constrained in combination with the strong radio background.

From the 1D posterior distribution, we see that the data constrain the radio production efficiency of the early sources, with the values of $L_\mathrm{r}/\mathrm{SFR} \gtrsim 1.549 \times 10^{25}$~W Hz$^{-1}$M$_\odot^{-1}$ yr at 150~MHz ($f_\mathrm{radio} \geq 1549$) being disfavoured at 68\% confidence. Moreover, we expect a significant correlation between the impact of radio background and that of the thermal history on the global 21-cm signal, as both a strong radio emission and weak X-ray heating contribute in the same direction deepening the absorption trough. Considering the 2D posterior probability in the plane $L_\mathrm{r}/\mathrm{SFR}-L_\mathrm{X}/\mathrm{SFR}$ and the corresponding approximate contours in red, we see that high-redshift galaxies that are both efficient in producing radio photons 
with $L_\mathrm{r}/\mathrm{SFR} \gtrsim 1.00\times10^{25}$~W Hz$^{-1}$M$_\odot^{-1}$ yr 
and inefficient at producing X-ray photons 
with an X-ray luminosity per star formation rate of $L_\mathrm{X}/\mathrm{SFR} \lesssim 1.09\times10^{42}\textnormal{erg~s}^{-1}$M$_\odot^{-1}$ yr
are disfavoured. $L_{r}/$SFR and $L_{X}/$SFR are proportional to $f_\mathrm{radio}$ and $f_X$ respectively and defined in \textit{Methods}.

The data disfavour (at 68\% confidence) models with early onset of efficient star formation which is characterized by low values of $5.37 \lesssim V_c \lesssim 15.5$~kms$^{-1}$ (note that the lower limit of the prior range is $V_c = 4.2$~kms$^{-1}$), corresponding to small typical dark matter halos of $4.4\times10^{5}$~M$_\odot \lesssim M \lesssim 1.1\times10^{7}$~M$_\odot$ at $z = 20$ \citep[e.g.][]{Reis2020},  and high values of star formation efficiency $f_* \gtrsim 0.05$, interpreted as a large fraction of collapsed gas that turns into stars. Each one of these  criteria individually, as well as their combination, would guarantee efficient Ly-$\alpha$ coupling, resulting in a deep high-redshift absorption profile. Considering the 2D posterior distribution, we find, using the approximate red contours on \cref{fig:fradio_results}, that $f_* \gtrsim 0.03$
together with galaxies hosted in dark matter halos of masses $M \lesssim 8.53\times10^8$ M$_\odot$ at $z = 20$ ($V_c \lesssim 31$~km/s) are disfavoured. We also find that combinations of high $f_*$ (and low $V_c$) with both low X-ray efficiency and high radio efficiency are disfavoured. We note that when fitting the data with the phenomenological synchrotron radio background model, we disfavour similar combinations of $V_c$ and $f_*$ (see \textit{Supplementary Material}).

The derived constraints can be compared to the recently published constraints from SARAS2 \cite{SARAS2}, see \cref{tab:numbers} and  \cref{fig:saras2_saras3_comparison} in \textit{Supplementary Material}. SARAS2 probes a lower redshift range, $z = 7 -12$,  and, thus, is complementary to SARAS3, being more sensitive to the process of heating and ionization. However, the experiment has a comparatively low signal-to-noise ratio, meaning that any constraints derived from it are likely to be weaker. For example, one particular signal may have a magnitude lower than the SARAS3 noise floor but higher than the SARAS2 noise floor. This means that were that particular signal real, SARAS3 would have detected it but SARAS2 would not of, and hence, given the non-detection in the SARAS3 data, the corresponding combination of astrophysical parameters will produce a lower posterior probability for SARAS3, $\mathcal{P}(\theta|D_\mathrm{SARAS3}, M_\mathrm{SARAS3})$, than SARAS2, $\mathcal{P}(\theta|D_\mathrm{SARAS2}, M_\mathrm{SARAS2})$~(see \textit{Methods} for a discussion on the posterior probability). Previously, it was found that SARAS2 disfavours (at approximately 68\% confidence) early galaxies with X-ray luminosity of $L_\mathrm{X}/\mathrm{SFR} \lesssim 6.3\times10^{39}\textnormal{erg~s}^{-1}$M$_\odot^{-1}$ yr in combination with $L_\mathrm{r}/\mathrm{SFR} \gtrsim 4.07\times10^{24}$~W Hz$^{-1}$M$_\odot^{-1}$ yr \cite{SARAS2}. This corresponds to disfavouring $\approx$23\% of the available parameter space in the $L_\mathrm{X} - L_\mathrm{r}$ plane at approximately 68\% confidence compared to $\approx$32\% for SARAS3, although we note that both experiments disfavour slightly different regions of the parameter space. 

The same set of astrophysical models, used here, has recently been constrained with an upper limit on the 21-cm power spectrum measured by HERA  \cite{HERA}. HERA disfavours at 68\% confidence level values of $L_\mathrm{r}/\mathrm{SFR} \gtrsim 4\times10^{24}$~W Hz$^{-1}$M$_\odot^{-1}$ yr as well as $L_\mathrm{X}/\mathrm{SFR} \lesssim 7.6\times10^{39}\textnormal{erg~s}^{-1}$M$_\odot^{-1}$ yr. We find that SARAS3 provides a similar constraint in the plane $L_\mathrm{r}/\mathrm{SFR}$ and $L_\mathrm{X}/\mathrm{SFR}$ with a weaker limit on $L_\mathrm{r}/\mathrm{SFR}$ but a stronger limit on $L_\mathrm{X}/\mathrm{SFR}$ than HERA. Similarly, SARAS2 gives a comparable constraint in the $L_\mathrm{r}/\mathrm{SFR}- L_\mathrm{X}/\mathrm{SFR}$ plane. We note that both SARAS2 and HERA probe the 21-cm signal at much lower redshifts than SARAS3 
thus the experiments potentially probe different populations of sources. Moreover,  HERA constraints come from the limit on the 21-cm power spectrum, rather than the global signal. Constraints on the 21-cm power spectrum are also available from the MWA~\cite{Trott_mwa_2020, Ghara_MWA_2021} and LOFAR~\cite{Gehlot_lofar_2019, Mondal_LOFAR_2020, LOFAR2021} interferometers. However, these limits are slightly weaker than those from HERA.

In the context of verifying the EDGES Low-Band detection, we assess the constraining power of the SARAS3 data on physical models that could, in principle, describe the reported absorption feature. Although, we note that none of our models can fit the flattened EDGES absorption signal well. We define EDGES-like signals using a conditional equation that ensures the models have approximately the same central frequency, width and depth as the EDGES absorption feature but does not strictly enforce the flattened Gaussian shape of the EDGES profile \cite{Fialkov2019, Reis_sta_2021}. In our analysis so far, the broad prior range was determined by our poor understanding of the high-redshift Universe. Now we use the restricted EDGES-like space as our prior, which is shown in \cref{fig:EDGES-like}.
We perform the fitting procedure and penalize models that do not meet the EDGES-like criteria by setting the likelihood to zero. The volume contraction from prior to posterior gives a quantitative measure of the level of consistency between the EDGES-like prior and the SARAS3 data and can be estimated using a marginal KL divergence \cite{kullback_information_1951}. This effectively allows us to say that if EDGES is true and indicative of a physical scenario, then a given percentage of the physical EDGES-like parameters space is inconsistent or ruled out by the SARAS3 data. We find that the volume of the EDGES-like posterior, when fitting with the radio galaxy models, is 60\% of the EDGES-like prior volume. In other words, 60\% of the physical EDGES-like parameter space is consistent with the SARAS3 data. See \textit{Methods} and \textit{Supplementary Material} for details.

Finally, we find that the data provide interesting limits on the amplitude of the synchrotron radio background in excess of the CMB disfavouring contributions  of $\gtrsim6$\% at a reference frequency of 1.42~GHz with 68\% confidence. The constraints from SARAS3 can be compared to the excess backgrounds inferred from ARCADE2 \cite{fixsen_arcade_2011} and LWA \cite{dowell_radio_2018} experiments, assuming that the excess is cosmological and is not due to incorrect calibration of the Galactic foregrounds \cite{Subrahmanyan2013}. We find that the 68\% confidence limit on $T_\mathrm{rad}$ is significantly lower than the reported deductions from the two experiments (see \textit{Supplementary Material}).

\section{Conclusion}\label{sec:conclusions}

We provide astrophysical constraints on the Universe at $z\sim 20$, corresponding to $\sim 200$ million years after the Big Bang, using upper limits on the sky-averaged 21-cm signal measured by the SARAS3 radiometer in the frequency range 55 and 85 MHz, $z\sim 15-25$. These are the first astrophysical limits of their kind. The only other existing constraining data (from EDGES) revealed a controversial flattened absorption profile, which is awaiting verification by an independent experiment. The residuals observed in SARAS3 data, after modelling for foregrounds, do not provide evidence for a detected 21-cm signal, including the EDGES profile, and they allow for the first time constraints of astrophysics at cosmic dawn.

We fit the data  with a log-log polynomial foreground model, as in the original SARAS3 data analysis paper, together with astrophysically motivated models for the global 21-cm signal, showing that deep global signals are disfavoured by the data. These constraints are then mapped into the astrophysical parameter space using a fully Bayesian analysis. We find that the SARAS3 data provide constraints on the processes that are linked to the formation of first stars and galaxies, production of radio photons at high redshifts as well as heating of the intergalactic medium.
We disfavour, at 68\% confidence, a population of radio galaxies with luminosity per SFR of $L_\mathrm{r}/\mathrm{SFR} \gtrsim 1.549 \times 10^{25}$~W Hz$^{-1}$M$_\odot^{-1}$ yr at 150~MHz, i.e. a factor of a thousand brighter than their low-redshift counterparts, and a synchrotron radio background in excess of the CMB of  $\gtrsim 6\%$ at 1.42 GHz. We also find correlation between the constraints on the radio background and on the thermal history of the global 21-cm signal, showing that galaxies which are both luminous in the radio band and inefficient at producing X-ray photons are disfavoured. Finally, the non-detection of the 21-cm signal in the SARAS3 data can be used to derive constraints on the properties of the first star forming regions. We find that, as an approximation to the 68\% confidence constraint, the data disfavour efficient star formation at high redshifts with a minimum mass of star forming halos of $M \lesssim 8.53\times10^8$~M$_\odot$ at $z=20$ in which $\gtrsim 3\%$ of the gas is converted into stars. 

Lessons learned from the SARAS3 analysis can be contrasted with those from other instruments, specifically with the EDGES Low-Band detection at $z\sim17$ as well as the astrophysical limits derived from the SARAS2 data at $z\sim7-12$ and the limits from HERA on the 21-cm power spectrum at $z\sim8-10$, MWA at $z\approx 6.7 - 8.5$ \cite{Trott_mwa_2020, Ghara_MWA_2021} and LOFAR at $z\approx 9$ \cite{Gehlot_lofar_2019, LOFAR2021}. For example, by conditioning the prior parameter space to be compatible with the EDGES detection and neglecting the steep walls of the feature, we find that $\sim60\%$ of the available parameter space is still consistent with the SARAS3 data. 

Although the SARAS3 constraints on the  high-redshift astrophysical processes are weak, the analysis presented here demonstrates the potential of the 21-cm line as a probe for the early Universe. The cosmic dawn constraints are expected to tighten in the next few years as the new low-frequency 21-cm experiments are coming online \cite{Zarka_nenuFar_2018, REACH_pipeline_2021, de_lera_acedo_reach_2022}.

\setcounter{section}{0}
\renewcommand\thesubsection{M.\arabic{subsection}}
\renewcommand\thesection{M}
\section{Methods} \label{sec:method}

\subsection{Nested Sampling}

To identify constraints on the parameter space of the global signal, we use the nested sampling \cite{skilling_nested_2004} algorithm implemented with \textsc{polychord} \cite{Handley2015a, Handley2015b}. Samples of the parameter space of a model, $M$, are derived using Bayes theorem
\begin{equation}
    P(\theta|D, M) = \frac{\mathcal{L}(\theta)\pi(\theta)}{\mathcal{Z}},
    \label{eq:posterior}
\end{equation}
where $\theta$ is the vector of model parameters, $D$ represents the data, $\mathcal{L}$ is the likelihood representing the probability that we observe the data given the model, $\pi$ is the prior probability representing our knowledge of the parameter space before we perform any fitting and $\mathcal{Z}$ is the evidence which normalizes the posterior, $P(\theta|D, M)$. Nested sampling generates samples from the likelihood and prior probabilities to numerically approximate $\mathcal{Z}$ and effectively sample the posterior. 
A higher value of $\mathcal{Z}$ when fitting model $M_1$ to the data in comparison to when fitting model $M_2$ indicates a preference for the former. This means that the evidence can be used to determine if a signal is present in the data or not. For example, if $M_1$ comprises just a foreground model and $M_2$ includes both a foreground and signal model then $\mathcal{Z}_{M_1} > \mathcal{Z}_{M_2}$ means that we do not require a signal model to effectively describe the data.

\begin{figure}[h!]
    \centering
    \includegraphics{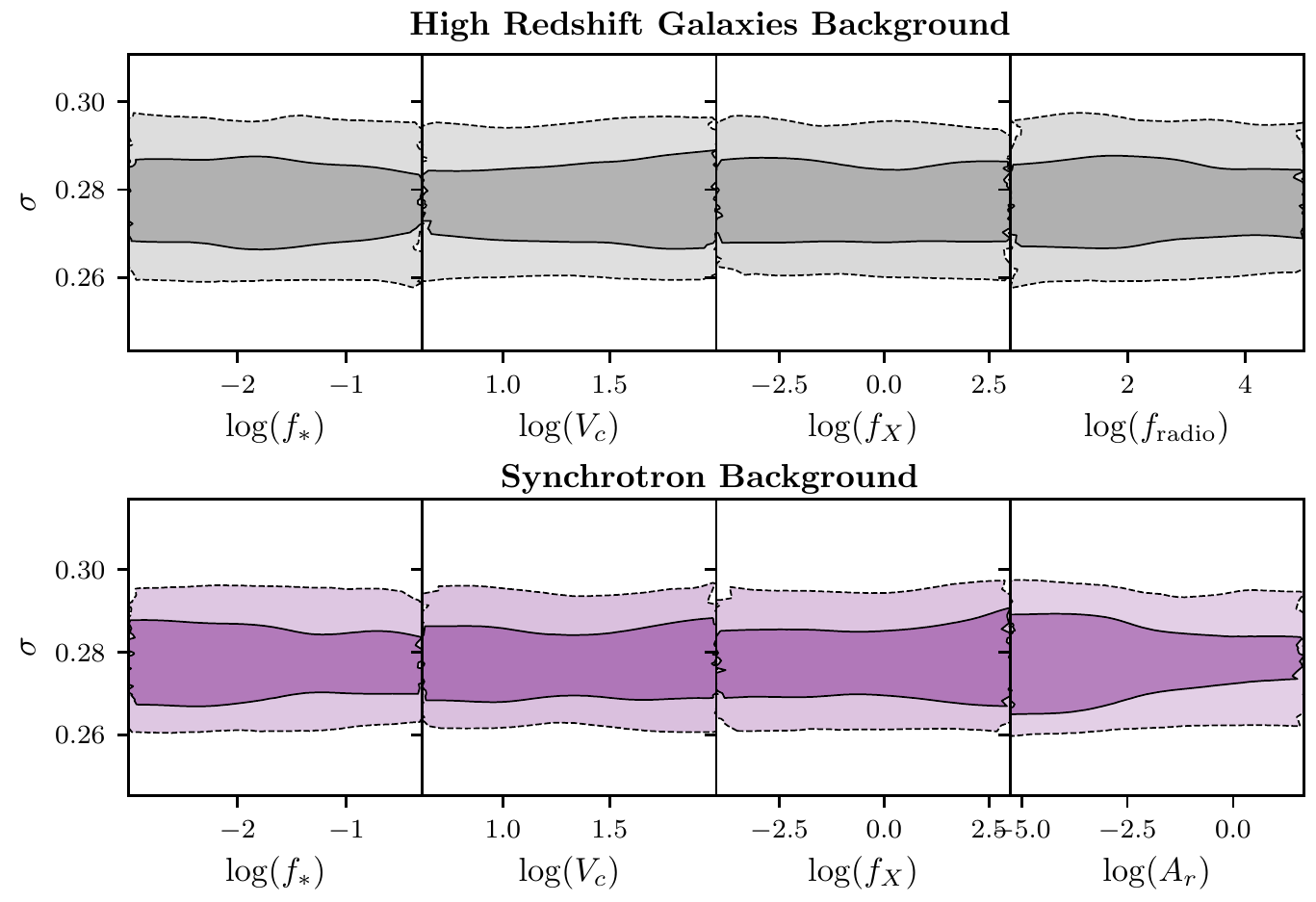}
    \caption{\textbf{The relationship between the astrophysical parameters and the noise.}}
    \label{fig:noise}
\end{figure}

The posterior distribution can then be interpreted as constraints on the model. The use of Bayesian inference is becoming more common in global 21-cm analysis and is an effective method to constrain  the astrophysical processes in the early Universe \cite{REACH_pipeline_2021, SARAS2, de_lera_acedo_reach_2022}. Throughout our analysis we assume a Gaussian likelihood function, $\mathcal{L}$, and a Gaussian noise distribution with a constant standard deviation, $\sigma$,
\begin{equation}
    \log\mathcal{L} = \sum_i \bigg(-\frac{1}{2}\log(2\pi\sigma^2) -\frac{1}{2}\bigg(\frac{T_\mathrm{D}(\nu_i) - T_\mathrm{fg}(\nu_i) - T_\mathrm{21}(\nu_i)}{\sigma}\bigg)^2\bigg),
    \label{eq:likelihood}
\end{equation}
where $T_\mathrm{D}$ is the SARAS3 data, $T_{21}$ is the global 21-cm signal model and $T_\mathrm{fg}$ is the foreground model. In practice, the noise in a global 21-cm experiment is expected to be larger at low frequencies and decrease with increasing frequencies, following the general trend of the sky temperature. A full treatment of any frequency dependence in the noise is left for future work. However, we find, see \cref{fig:noise}, that the posterior probability for the constant standard deviation on the assumed Gaussian noise is uncorrelated with the astrophysical parameters, and we would therefore expect a full treatment of the noise to have little impact on the derived parameter constraints for the two excess background models. We expect a full treatment of the noise to be more important in future experiments that provide tighter constraints on the astrophysical process during CD.

\subsection{Foreground Modelling}
The foreground model used here is identical to the one employed in the original SARAS3 analysis \cite{SARAS3}.  The log-log polynomial foreground model  is given by
\begin{equation}
    \log_{10}T_\mathrm{fg} = \sum_{i=0}^{i=6}a_i\left(\mathcal{R}(\log_{10}\nu)\right)^i,
\end{equation}
where $a_i$ are the fitted coefficients, $\mathcal{R}$ is a normalizing function that scales its argument, $\log_{10}\nu$, linearly between -1 and +1 and $\nu$ is frequency in MHz. When fitting the model with \textsc{polychord} we provide a uniform prior, our initial assumption about the model parameters, of $-10$ to 10 on each of the foreground model coefficients, $a_i$. In addition to the foregrounds, the model is designed to account for any residual systematics from the calibration process.

\subsection{Signal Modelling and Emulation}

At the high redshifts of CD the most important factors that drive the 21-cm signal are the intensity of the Ly-$\alpha$ background which determines the efficiency of the coupling between $T_\mathrm{S}$ and $T_\mathrm{K}$, the temperature of the radio background, $T_\mathrm{rad}$, and the thermal history of the gas. The dependence of the 21-cm signal on these processes is as follows: The earlier star formation starts, the lower will be the frequency of the absorption profile; the stronger the Ly-$\alpha$ background, the steeper and deeper will be the resulting 21-cm signal; the colder is the gas, relative to the background radiation, the deeper will be the absorption feature. The resulting 21-cm signal can be written as 
\begin{equation}    
T_{21} = \frac{T_\mathrm{S} - T_\mathrm{rad}}{1+z} \left[1 - \exp(-\tau_{21})\right] \propto 1-\frac{T_\mathrm{rad}}{T_\mathrm{S}},
    \label{eq:t21}
\end{equation}
where we assumed that the Universe is largely neutral at the high redshifts of CD.

One potential source of radio photons at CD are early radio galaxies \citep{Mirocha2019}. The radio background contribution created by such sources is proportional to the star formation rate~(SFR), thus increasing with time,  and  is non-uniform following the distribution of galaxies. The radio luminosity spectrum as a function of frequency $\nu$ produced by a star forming region and calculated per SFR in units of WHz$^{-1}$ is given by 
\begin{equation}
    L_\mathrm{r} = f_\mathrm{radio} 10^{22} \bigg(\frac{\nu}{150~\mathrm{MHz}}\bigg)^{-\alpha_\mathrm{radio}} \frac{\mathrm{SFR}}{\mathrm{M}_\odot\mathrm{yr}^{-1}},
    \label{eq:radio_luminosity}
\end{equation}
where $f_\mathrm{radio}$ is an efficiency factor that measures radio photon production in high-redshift galaxies compared to their present day counterparts and $\alpha_\mathrm{radio}=0.7$ is the spectral index in the radio band \cite{Reis2020}. The temperature of the radio background produced by such galaxies at redshift $z$ is calculated by integrating over the contribution of all galaxies within the past light-cone \cite{Reis2020} and is added to the temperature of the CMB to give the total radio background temperature. 
We quote constraints on the radio luminosity per SFR, $L_\mathrm{r}/\mathrm{SFR}$, at a reference frequency of 150~MHz in section \ref{sec:results}.

We take into account several heating and cooling mechanisms, such as cooling due to the expansion of the Universe and heating due to structure formation, Ly-$\alpha$ \cite{Madau, Chuzhoy2007} and CMB \cite{Venumadhav2018} heating, as well as heating by first X-ray binaries \cite{Fialkov2014Natur}. In our model, the first four effects are fully determined by cosmology and star formation, whereas heating by X-ray binaries invokes new astrophysical processes (such as black hole binary formation and X-ray production by the high redshift sources). Therefore, X-ray heating requires independent parameterization, and we model X-ray luminosity per SFR \cite{Fragos_Xrays_2013} as 
\begin{equation}
    L_\mathrm{X, 0.2 - 95 \textnormal{keV}} = 3\times10^{40} f_X \frac{\mathrm{SFR}}{\mathrm{M}_\odot\mathrm{yr}^{-1}}
\end{equation}
calculated in units of erg~s$^{-1}$ between 0.2 and 95 keV, where $f_X$ is the efficiency of X-ray photon production.
Gas thermal history is then evaluated at every redshift by integrating over the contribution of all galaxies within the past light-cone to find the corresponding heating rate and then solving a differential equation to evolve the gas temperature. 

Both the radio luminosity and the thermal history of the gas depend on the SFR, which is not well constrained for early galaxies. Therefore, our model also includes free parameters that regulate star formation. One is the star formation efficiency of high-redshift galaxies, $f_*$, which measures the fraction of collapsed gas in star forming regions that turns into stars, and the other is the minimum  mass of star forming halos, or, equivalently, the minimum circular velocity of star forming halos, $V_c$ \cite{Barkana_mass_2001}.
This quantity depends on the local environment of each star forming region and is affected by factors such as the local intensity of the radiative background in the Lyman-Werner band \cite{Fialkov2013, Schauer2021} or the relative velocity between dark matter and gas \cite{Tseliakhovich2010, Fialkov2012, Schauer2021}.

In order to physically model the global 21-cm signal, we rely on neural network-based emulation with the \textsc{python} package \textsc{globalemu} \cite{Bevins_globalemu_2021} trained on the results of the full semi-numerical simulations of the global 21-cm signal \citep[][]{Visbal2012, Fialkov2014, Fialkov2019, Cohen_charting_2017, Reis2020, Reis_sta_2021}.

For each global signal model, we have a series of testing and training signals. \Cref{tab:priors} shows the ranges of the parameters in each of the training and testing data sets for the different models of the global 21-cm signals used in this work. The boundaries correspond to the broadest possible ranges allowed for each one of the parameters from the astrophysical principles and existing (weak) observational constraints. Outside these ranges the emulators are unreliable and consequently the ranges act as the prior bounds for the nested sampling code \textsc{polychord}. The parameters are sampled either uniformly or log-uniformly between the ranges in the training and test data, and we use appropriate prior probability distributions for each parameter when running the fits.

For all three signal emulators, we use the same neural network architecture with 4 hidden layers of 16 nodes each. The same radio galaxies and CMB only emulators were recently used in our analysis of the SARAS2 data \cite{SARAS2}. We note that the network for the CMB only radio background models, however, has seven astrophysical inputs compared to the radio galaxy and synchrotron radio background networks which both have five. For the CMB only model the X-ray spectral energy density~(SED) is characterized by the slope of the spectrum, $\alpha$, and a low energy cut-off, $E_\mathrm{min}$; while  for the other two models the X-ray SED is fixed to that of high-redshift X-ray binaries \cite{Fragos_Xrays_2013}. Further, parameters related to reionization have very modest effect on the 21-cm signal in the SARAS3 range. Therefore, we fix the mean free path of ionizing photons. For the radio galaxies and radio synchrotron backgrounds, the mean free path was fixed to 40~Mpc, while in the CMB-only case it was fixed to  $R_\mathrm{mfp} = 30$~Mpc. 

We assess the accuracy of the neural networks in the SARAS3 band, $z \approx 15 - 25$, using the root mean squared error~(RMSE) when emulating the test data after training. The synchrotron radio background network has a mean RMSE when emulating 1034 test models, after training on 9304 models, of 7.98~mK, a 95$\textsuperscript{th}$ percentile RMSE of 23.06~mK and a worst RMSE of 85.65~mK. The CMB only background network is trained on 5137 models and tested on 570 models. In the SARAS3 band the mean RMSE for the test data is 0.78~mK, the 95$\textsuperscript{th}$ percentile is 2.67~mK and the worst is 13.36~mK. Finally, when trained on a data set of 4311 models and tested on 479 models the radio galaxy radio background neural network emulator is found to have a mean RMSE of 5.11~mK, a 95$\textsuperscript{th}$ percentile RMSE of 20.53~mK and a worst RMSE of 81.70~mK in the SARAS3 band. All the trained networks have 95$\textsuperscript{th}$ percentile RMSEs well below the RMS found after modelling and subtracting the log-log polynomial foreground model from the SARAS3 data. The numbers are summarized in \cref{tab:networks}.

\begin{figure}[h!]
    \centering
    \includegraphics{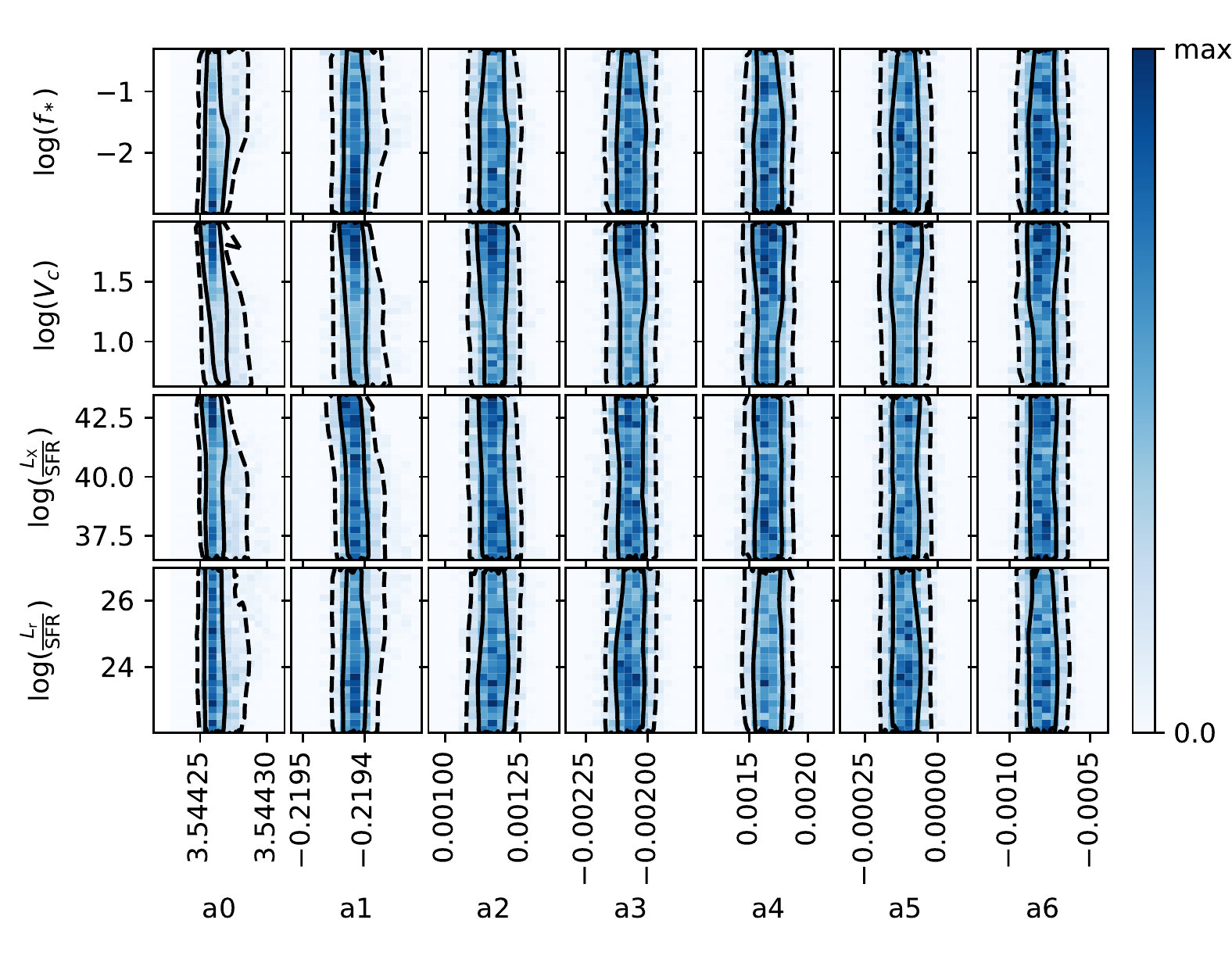}
    \caption{\textbf{The 2D posterior distributions between the astrophysical and foreground parameters found when fitting the data with the radio galaxy radio background models.}}
    \label{fig:correlations}
\end{figure}

When fitting all three signal models with the foreground, we see no correlation between the astrophysical parameters and the foreground parameters. An example of this can be seen in the 2D posteriors, which are shown in \cref{fig:correlations}, between the astrophysical and foreground parameters from the fit with the radio galaxies background model.

\subsection{Marginal KL Divergence and EDGES-like signals}

To determine the volume of the EDGES-like parameter space that the SARAS3 data rules out, we calculate a marginal Kullback-Leibler~(KL) divergence, $\mathcal{D}$. To illustrate the types of signals that we are selecting by constraining our parameter space to be EDGES-like, we show the corresponding functional prior and posterior in \textit{Supplementary Material}.

The KL divergence is a measure of the information gained when contracting a prior onto a posterior. For our purposes we are interested in the EDGES-like parameter space and as a result we would consider the foreground parameters to be nuisance parameters and need to integrate them out.

To calculate the marginal KL divergence, $\mathcal{D}$, we therefore need to evaluate the log-probabilities associated with the signal parameters in the EDGES-like prior, $\pi$, and the corresponding posterior, $\mathcal{P}$,
\begin{equation}
    \mathcal{D(\mathcal{P}||\mathcal{\pi})} = \int \mathcal{P}(\theta) \log_{e}\bigg(\frac{\mathcal{P}(\theta)}{\mathcal{\pi}(\theta)}\bigg) d\theta = \bigg\langle \log_{e}\bigg(\frac{\mathcal{P}(\theta)}{\mathcal{\pi}(\theta)}\bigg) \bigg\rangle_\mathcal{P}.
    \label{eq:kl_divergence}
\end{equation}
We use a Gaussian kernel density estimators~(KDE) to replicate the samples in the signal sub-spaces of our EDGES-like prior and posterior via the recently developed code \textsc{margarine}
\cite{margarine_neurips, margarine_maxent}. A multivariate Gaussian KDE, implemented with \textsc{scipy}, is produced by summing over multiple multivariate Gaussian profiles with known standard deviation centred around each sample point, consequently the log of the probability density function is easily tractable making the KL divergence easily calculable. 

When training our KDEs on the prior and posterior samples, we first transform our data into the standard normal parameter space, which improves the accuracy of the density estimator and allows it to better capture the sharp edges of approximately flat distributions.

It can be shown that the KL divergence is related to the volume fraction between the posterior and the prior via
\begin{equation}
    \mathcal{D}(\mathcal{P}||\mathcal{\pi}) = \log_{e}\bigg(\frac{V_\mathcal{P}}{V_\pi}\bigg),
\end{equation}
and therefore
\begin{equation}
    \exp(-\mathcal{D}(\mathcal{P}||\mathcal{\pi})) = \frac{V_\pi}{V_\mathcal{P}},
\end{equation}
can be used to determine the volume of the prior contained in the posterior or in our case the volume of the EDGES-like prior that is still consistent with the SARAS3 data after fitting.

\section*{Acknowledgments}

The authors would like to thank the reviewers for their helpful comments regarding our manuscript. HTJB acknowledges the support of the Science and Technology Facilities Council (STFC) through grant number ST/T505997/1. WJH and AF were supported by Royal Society University Research Fellowships. EdLA was supported by the STFC through the Ernest Rutherford Fellowship. RB acknowledges the support of the Israel Science Foundation (grant No. 2359/20), The Ambrose Monell Foundation and the Institute for Advanced Study as well as the Vera Rubin Presidential Chair in Astronomy and the Packard Foundation.

\section*{Author Contribution}

HTJB performed the data analysis and led the writing of the paper. AF initiated the project, supervised it  and helped writing and revising the article. EdLA supervised the project and the analysis, and helped with the writing and revision of the article. WJH provided technical support and advice regarding the Bayesian methodology. The analysis in the paper is of non-public data provided by RS and SS. The astrophysical signal models were provided by AF and RB. All co-authors provided comments and contributed to the structure of the article.

\section*{Competing Interests}

The authors declare that they have no competing interests.

\section*{Data Availability}

The SARAS3 data are available upon reasonable request to SS.

\section*{Code Availability}

\textsc{globalemu} is available at \url{https://github.com/htjb/globalemu} and \textsc{margarine} at \url{https://github.com/htjb/margarine}. The nested sampling tool \textsc{polchord} at \url{https://github.com/PolyChord/PolyChordLite} and the nested sampling post-processing codes, \textsc{anesthetic} and \textsc{fgivenx}, are available at \url{https://github.com/williamjameshandley/anesthetic} and \url{https://github.com/williamjameshandley/fgivenx} respectively. All other codes used are available upon reasonable request to HTJB.
\\

\section*{Tables}

\bgroup
\def\arraystretch{1.5}
\begin{table}[h!]
    \centering
    \begin{tabular}{|c|c|c|c|}
    \hline
         Background & Radio Galaxies & Synchrotron & CMB Only \\
         \hline
         \hline
         Training Models & 4311 & 9304 & 5137 \\
         \hline
         Testing Models & 479 & 1034 & 570 \\
         \hline
         \hline
         Mean RMSE & 5.11 & 7.98 & 0.78 \\
         \hline
         95 Percentile RMSE & 20.53 & 23.06 & 2.67 \\
         \hline
         Worst RMSE & 81.70 & 85.65 & 13.36\\
         \hline
    \end{tabular}
    \caption{\textbf{The neural network emulation.} The table summarizes the number of models used to train and test the three different neural network emulators used in this paper, along with summary statistics for their accuracies. Temperatures RMSEs are given in mK.}
    \label{tab:networks}
\end{table}
\egroup

\bgroup
\def\arraystretch{1.5}
\begin{table}[h!]
\centering
\begin{tabular}{|c|c|c|}
\hline
Parameter & Radio Background & Range \\
\hline
\hline
$f_*$ & CMB Only, Synchrotron, Radio Galaxies & 0.001 - 0.5 \\
\hline
$V_c$ & CMB Only, Synchrotron, Radio Galaxies & 4.2 - 100 km/s\\
\hline
$f_X$ & CMB Only, Synchrotron, Radio Galaxies & 0.001 - 1000\\
\hline
$f_\mathrm{radio}$ & Radio Galaxies & 1.0 - 99,500 \\
\hline
$A_{\mathrm{r}}^{1420}$ & Synchrotron & 0 - 47 \\
\hline
\multirow{3}*{$\tau$} & CMB Only & 0.026 - 0.103\\\cline{2-3}
& Synchrotron & 0.016 - 0.158\\\cline{2-3}
& Radio Galaxies & 0.035 - 0.077\\
\hline
$\alpha$ & CMB Only & 1.0 - 1.5 \\
\hline
$E_\mathrm{min}$ & CMB Only & 0.1 - 3.0 keV\\
\hline
$R_\mathrm{mfp}$ & CMB Only, Synchrotron, Radio Galaxies & Fixed at 30, 40 and 40 ~Mpc\\
\hline
\end{tabular}
\caption{\textbf{The astrophysical priors.} The prior ranges on the parameters for the CMB only, synchrotron and high-redshift radio galaxy background global 21-cm signal models fitted in this paper. The definitions of the parameters are given in the text. The prior ranges are designed to encompass the current uncertainty in the properties of the high-redshift Universe. The emulators are unreliable outside these bounds. Note that $\tau$ is not an important parameter in the SARAS3 band, however, we train the models with this parameter as an input, perform fits with it and then marginalize over it. Similarly, $R_\mathrm{mfp}$ is only important at lower redshifts outside the SARAS3 band. The global signal only has a weak dependence on this parameter, and so we fix its value of 40~Mpc for the radio galaxies and radio synchrotron backgrounds, while in the CMB-only case it was fixed to  $R_\mathrm{mfp} = 30$~Mpc. }
\label{tab:priors}
\end{table}
\egroup

\bgroup
\def\arraystretch{1.5}
\begin{table}[h!]
    \centering
    \begin{tabular}{|c|c|c|c|}
         \hline
         & SARAS3 & HERA & SARAS2 \\
         \hline
         \hline
         Signal type & Global & Power Spectrum & Global \\
         \hline
         Redshift range & $z\approx 15 - 25$ & $z\approx 8$ and $\approx 10$ & $z\approx 7 - 12$\\
         \hline
         \hline
         $L_{r}/\mathrm{SFR}$ & $\gtrsim 1.549\times10^{25}$ & $\gtrsim4.00\times10^{24}$ & -- \\
         \hline
         $L_{r}/\mathrm{SFR} \cap L_{X}/\mathrm{SFR}$ & $\gtrsim 1\times10^{25} \cap \lesssim 1.09\times10^{42}$ & $\gtrsim 4.00\times10^{24} \cap \lesssim 7.60\times10^{39}$ & $\gtrsim 4.07\times10^{24} \cap \lesssim 6.3\times10^{39}$\\
         \hline
         $M$ & $4.4\times10^{5} \lesssim M \lesssim 1.1\times10^{7}$ & -- & --\\
         \hline
         $f_*$ & $\gtrsim 0.05$ & -- & --\\
         \hline
         $f_* \cap M$ & $\gtrsim 0.03 \cap \lesssim 8.53\times10^{8}$ & -- & --\\
         \hline
    \end{tabular}
    \caption{\textbf{Summary of key constraints from SARAS3 (this work), HERA \cite{HERA} and SARAS2 \cite{SARAS2} experiments.} We specify the signal type measured by each instrument (either  global signal or power spectrum); redshift range targeted by each experiment; constraints on the value of $L_r$/SFR expressed in units of  W Hz$^{-1}$M$_\odot^{-1}$ yr at 150~MHz; limits on $L_r$/SFR in combination with $L_X/$SFR (calculated between 0.2 and 95 keV and  expressed in units of  $\textnormal{erg~s}^{-1}$M$_\odot^{-1}$ yr); limits on  the mass of star forming halos,  $M$, given in solar masses at $z=20$, star formation efficiency $f_*$ and, finally, constraint on $f_*$ in combination with the halo mass. Limits on the individual parameters correspond to the regions that are disfavoured (with 68\% confidence) in the 1D posteriors, combined constraints approximately correspond to the 68\% confidence limits  in the 2D posteriors. Note that: SARAS2 is unable to constrain individual parameters;  HERA targets the power spectrum in comparison to the two SARAS experiments which attempt to measure the sky-averaged signal; SARAS3 is at much higher redshifts than the other two experiments; while HERA provides individual constraints on  $L_r$/SFR and $L_X/$SFR \cite{HERA}, here we only quote the  individual constraint on $L_r/$SFR and the combined constraint with $L_X/$SFR, which is done to ease the comparison with SARAS3 and SARAS2.}
    \label{tab:numbers}
\end{table}
\egroup

\pagebreak

\section*{Figure Legends/Captions (main text)}

\textbf{Figure 1: SARAS3 constraints on high-redshift radio galaxies.} The data disfavour deep global signals, as can be seen by comparing the functional prior (blue) with the posterior (red) in panel (a). At the bottom of this panel we show the Kullback-Leibler (KL) divergence, $\mathcal{D}$, as a function of redshift between the functional prior and posterior. The KL divergence gives a measure of the information gain when moving from one to the other and illustrates the constraining power of the SARAS3 data, which peaks at around $z\approx20$. Panel (b) shows the corresponding 1D and 2D posteriors for the astrophysical parameters found when fitting the foreground and a global 21-cm signal. From left to right (see text for details): fraction of gas that turns into stars, $f_*$;  circular velocity in units of km s$^{-1}$, $V_c$; radio luminosity per unit SFR, $L_\mathrm{r}/\mathrm{SFR}$, in units of ~W Hz$^{-1}$M$_\odot^{-1}$ yr calculated at 150~MHz; X-ray luminosity per unit SFR, $L_\mathrm{X}/\mathrm{SFR}$,  in $\textnormal{erg~s}^{-1}$M$_\odot^{-1}$ yr. The corresponding 1D posterior with marked 68\% disfavoured regions are shown in the top of each column. The colour of the 2D posteriors (see colour bar) reflects the magnitude of the 2D posterior probabilities. The dashed black lines encapsulate the 95\% confidence regions. The solid black lines show the 68\% confidence regions for which we make a conservative approximation with the dashed red lines (to guide the eye) with the corresponding numerical values summarized in the inverted triangle table. Figures produced with \textsc{anesthetic} \protect\cite{anesthetic} and \textsc{fgivenx} \protect\cite{fgivenx}.

\textbf{Figure 2: The relationship between the astrophysical parameters and the noise.} The figure shows that the standard deviation of the assumed Gaussian noise is uncorrelated with the astrophysical parameters, and consequently we would expect that a full treatment of any frequency dependence of the noise in the data, which is left for future work, will have little impact on the derived parameter constraints.

\textbf{Figure 3: The 2D posterior distributions between the astrophysical and foreground parameters found when fitting the data with the radio galaxy radio background models.} We see no clear correlations between the two sets of parameters, indicating that they are independent of each other.

\pagebreak

\bibliographystyle{naturemag}
\bibliography{journals,ref}
\pagebreak

\pagebreak


\setcounter{section}{0}
\renewcommand\thesubsection{S.\arabic{subsection}}
\renewcommand\thesection{S}

\section{Supplementary Material} \label{sec:sup_mat}



\subsection{Foreground Modelling: Comparison With Previous Work}

\begin{figure}[h!]
    \centering
    \includegraphics{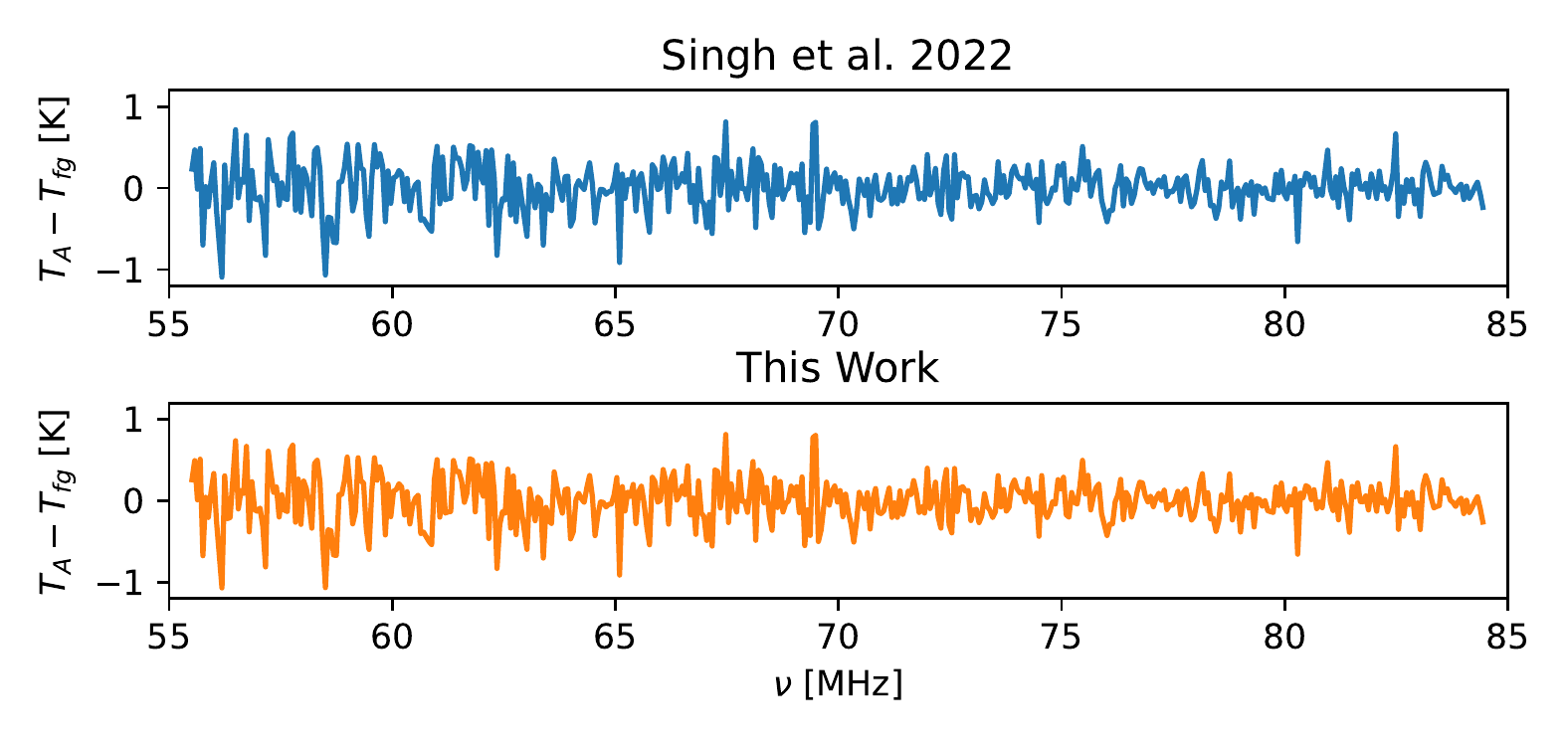}
    \caption{\textbf{A comparison of foreground models used here and in the original analysis.} The figure shows the residuals found when subtracting the foreground model parameterized by the MCMC run in the original SARAS3 analysis in comparison to the equivalent using the parameterization from our nested sampling run. We see that the two sets of residuals are consistent with each other, showing that our implementation of the foreground is consistent with the previously implemented MCMC run.}
    \label{fig:fore_comparison}
\end{figure}

In \cref{fig:fore_comparison}, we show the residuals found when subtracting the foreground model parameterized by the nested sampling results in this work and parameterized by the MCMC run in the original work \cite{SARAS3}. The two sets of residuals are found to be consistent with each other, indicating that our implementation of the foreground modelling is also consistent with that in the original SARAS3 analysis.

\subsection{Comparison of Constraints on the High-redshift Radio Galaxy Model from SARAS2 and SARAS3}
\label{sm:saras2}

\begin{figure}[h!]
    \centering
    \includegraphics{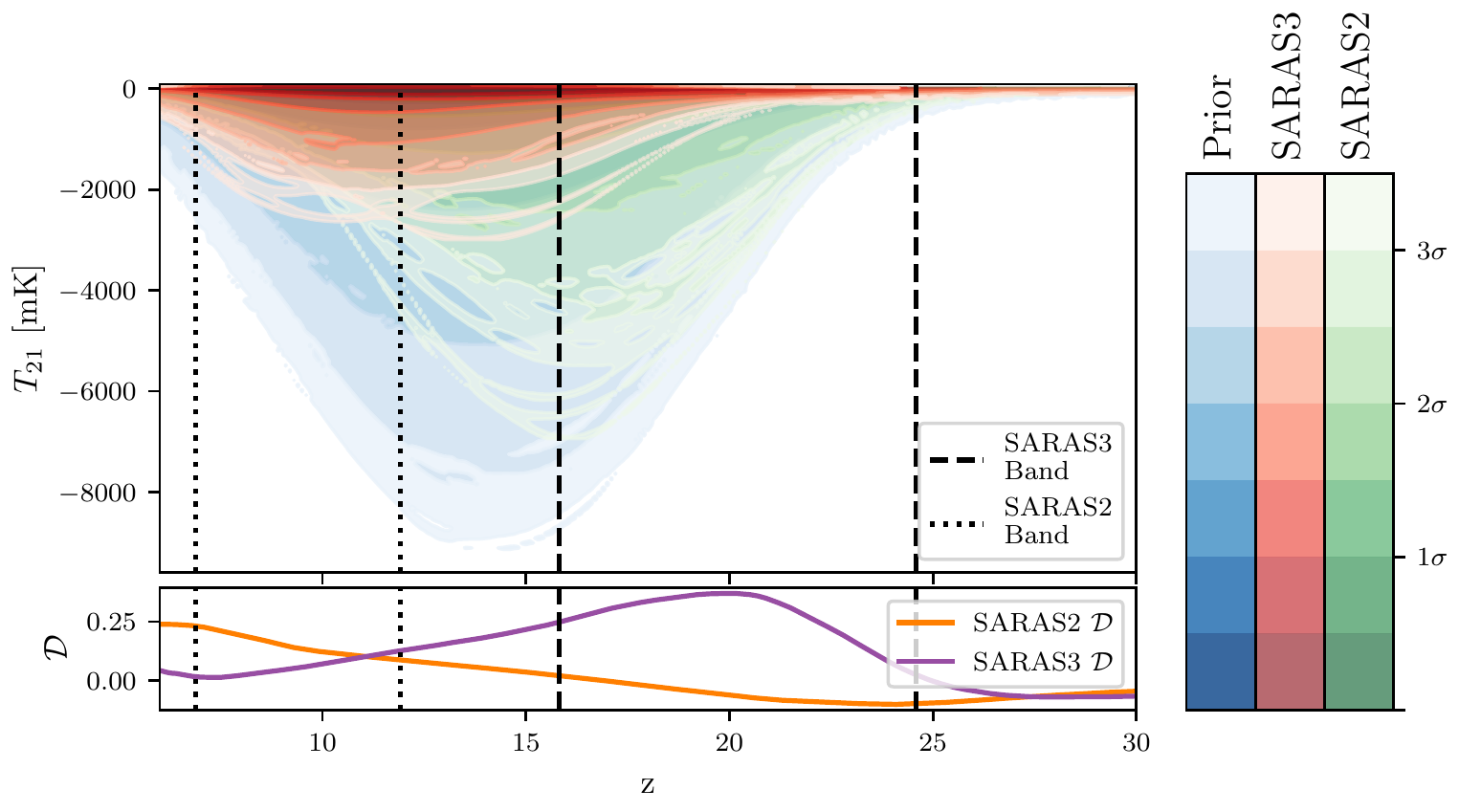}
    \caption{\textbf{A comparison of the constraints from the SARAS2 analysis \cite{SARAS2} and the SARAS3 analysis presented here when fitting both data sets with the radio galaxy background models.} Note that the two sets of analysis have a number of differences, and so direct comparisons like this have to be made with caution. Namely, the SARAS2 functional posterior is derived from a number of fits to allow for uncertainty in the modelling of the noise and a systematic,  whereas the SARAS3 results are derived from a single fit of a foreground plus signal model. Both sets of analysis share the same prior. Nonetheless, we can see from the KL divergence in the bottom panel that the SARAS3 data provides tighter constraints on this astrophysical models at $z \gtrsim 10$ and the SARAS2 data at $z \lesssim 10$.}
    \label{fig:saras2_saras3_comparison}
\end{figure}

We note that, previous analysis of the SARAS2 data used similar techniques to those detailed in this paper and fitted the SARAS2 data, in the range $z \approx 7 -12$, with the radio galaxy  global signal models, maximally smooth foreground models and models for non-smooth systematics \cite{SARAS2}. Although the signal-to-noise ratio is lower for the SARAS2 data, the analysis disfavoured the deepest signals corresponding to the  combinations of parameters: high $f_\mathrm{radio}$ with high $f_*$ and low $f_X$. Comparison of the functional posterior plots, shown in \cref{fig:saras2_saras3_comparison}, demonstrates that the two experiments are complimentary, with the peak constraining power of  SARAS3  at $z \gtrsim 10$ and that of SARAS2 at  $z \lesssim 10$.

Looking at the 2D posterior distributions of the astrophysical parameters, we see that SARAS2  disfavours (at approximately 68\% confidence) galaxies with $f_X \lesssim 0.21$ ($L_\mathrm{X}/\mathrm{SFR} \lesssim 6.3\times10^{39}\textnormal{erg~s}^{-1}$M$_\odot^{-1}$ yr) in combination with $f_\mathrm{radio} \gtrsim 407$ ($L_\mathrm{r}/\mathrm{SFR} \gtrsim 4.07\times10^{24}$~W Hz$^{-1}$M$_\odot^{-1}$ yr). SARAS3  constrains a similar combination of parameters, with $f_\mathrm{radio} \gtrsim 1000$ ($L_\mathrm{r}/\mathrm{SFR} \gtrsim 1.00\times10^{25}$~W Hz$^{-1}$M$_\odot^{-1}$ yr) and $f_X \lesssim 109$ ($L_\mathrm{X}/\mathrm{SFR} \lesssim 1.09\times10^{42}\textnormal{erg~s}^{-1}$M$_\odot^{-1}$ yr) being disfavoured at $\approx68$\% confidence. SARAS2 was also found to disfavour galaxies with a combination of $f_* \gtrsim 0.03$ and  $f_\mathrm{radio} \gtrsim 707$ ($L_\mathrm{r}/\mathrm{SFR} \gtrsim 7.07\times10^{24}$~W Hz$^{-1}$M$_\odot^{-1}$ yr at 150~MHz) at $\approx68$\% confidence which is again comparable to the constraints from SARAS3, $f_* \gtrsim 0.03$ and $f_\mathrm{radio} \gtrsim 167$ ($L_\mathrm{r}/\mathrm{SFR} \gtrsim 1.67\times10^{24}$~W Hz$^{-1}$M$_\odot^{-1}$ yr at 150~MHz), except that SARAS3 again provides a tighter constraint on the radio luminosity. A major difference between the constraining power of the two experiments is that SARAS3 provides constraints on the individual parameters (as can be seen from the 1D posterior distributions) which SARAS2 is not able to do. The constraints of SARAS2 are manifested in higher-dimensional  posterior distributions, e.g. in 2D.

\subsection{Synchrotron Radio Background Constraints with SARAS3}
\label{synchRB}

In addition to our main model, where the excess radio background over the CMB is created by high-redshift radio galaxies, we consider a phenomenological synchrotron radio background \cite{Fialkov2019}. The total temperature of the radio background is 
\begin{equation}
    T_\mathrm{rad} = T_\mathrm{CMB}\bigg[ 1 + A_{\mathrm{r}}^{78}\bigg(\frac{\nu}{78\textnormal{MHz}}\bigg)^\beta\bigg],
    \label{eq:srb_background}
\end{equation}
where $T_\mathrm{CMB}$ is the CMB contribution, and we add the synchrotron term with amplitude $A_{\mathrm{r}}^{78}$ calculated at the reference frequency of 78 MHz. $\beta = -2.6$ is the power law index  which is chosen  in agreement with the observations of the radio background today by LWA \cite{dowell_radio_2018} and ARCADE2 \cite{fixsen_arcade_2011} low frequency radio experiments. Throughout this work we  also use the value of the amplitude at 1.42~GHz, $A_{\mathrm{r}}^{1420}$, and the relationship between the two amplitudes is given as
\begin{equation}
    A_{\mathrm{r}}^{78} = A_{\mathrm{r}}^{1420} \bigg(\frac{0.078}{1.420}\bigg)^\beta,
\end{equation}
where the superscript gives the reference frequency in MHz. 

\begin{figure}[h!]
    \centering
    \includegraphics{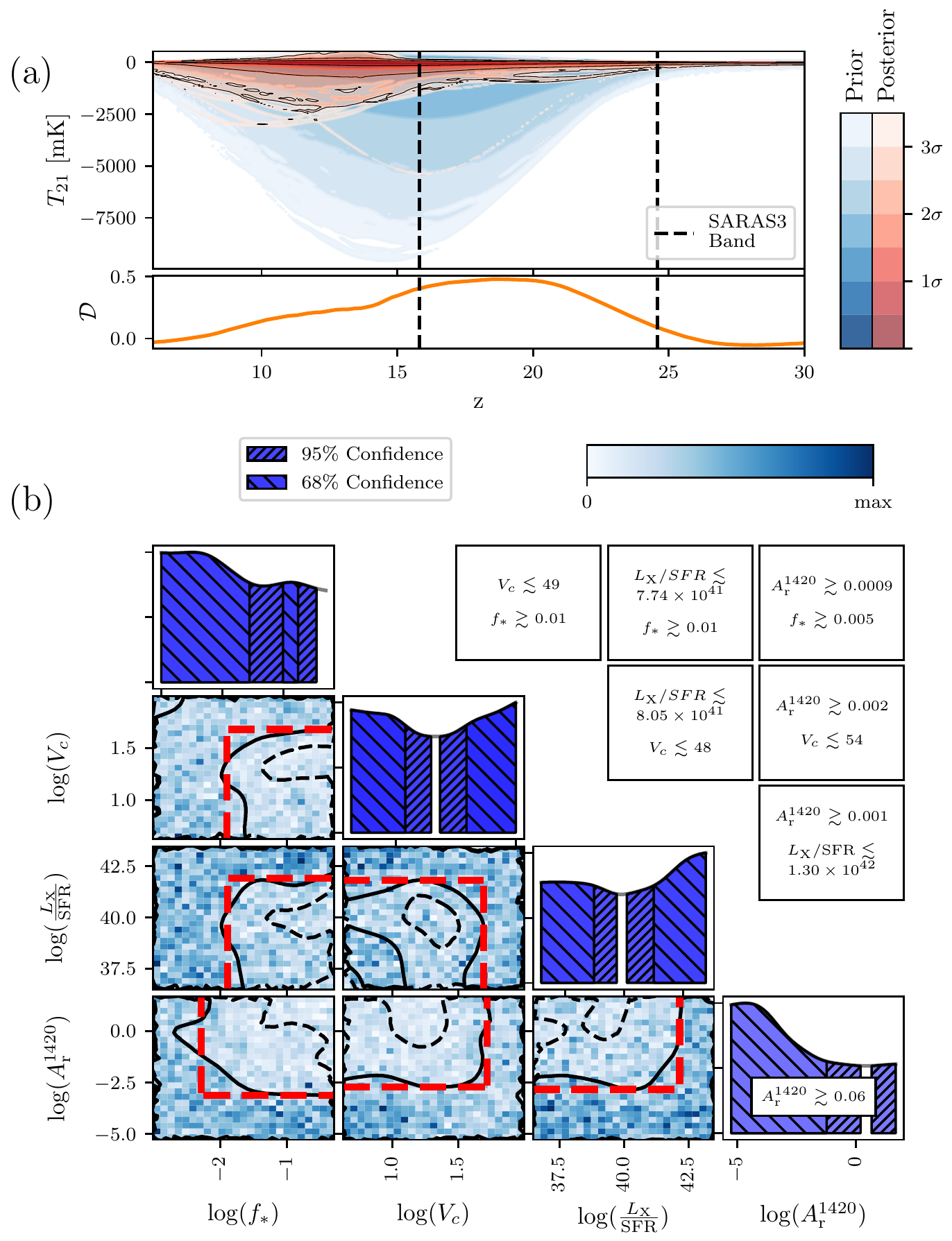}
    \caption{\textbf{SARAS3 constraints on early Universe scenarios with a synchrotron radio-background.} We disfavour the deepest synchrotron background signals, and the contraction from the functional prior (blue) to the functional posterior (red) is shown in panel (a) of this figure. We see that the KL divergence $D\approx0$ at high redshifts as expected and that the divergence is highest in the SARAS3 band around $z\approx20$.
    Panel (b) shows that the parameter $A_{\mathrm{r}}^{1420}$ is tightly constrained to $\lesssim 0.06$ at 68\% confidence and at a reference frequency of 1.42~GHz, corresponding to a radio background in excess of the CMB by $\lesssim 6\%$. High values of $f_*$, low values of $V_c$ in combination with high $A_\mathrm{r}$ and similarly low values of $L_\mathrm{X}/\mathrm{SFR}$ in combination with high values of $A_\mathrm{r}$ are all disfavoured. Again we show the approximate 2D 68\% disfavoured regions in the inverted corner plot and the 68\% 1D constraints in their respective sub-panels.}
    \label{fig:ARad_results}
\end{figure}

The two main differences between the radio background created by galaxies and the synchrotron radio background case is that the former is non-uniform and grows with time, following galaxy formation. In contrast, the latter is assumed to be uniform and decays with time, by construction. The different evolution of the two types of backgrounds is the key feature that leads to distinct astrophysical constraints with SARAS3.  We repeat our analysis for the synchrotron radio background model and show the results in  \cref{fig:ARad_results}.

Panel (a) of \cref{fig:ARad_results} shows the functional prior~(blue) for the synchrotron radio background model in comparison to the functional posterior~(red) corresponding to the constraints in panel (b). We see that, as with the radio galaxies background model, the SARAS3 data disfavours the deepest astrophysical models in the band $z \approx 15 -25$ but allows for some deeper models with minima at $z \lesssim 15$. The corresponding KL divergence, $\mathcal{D}$, between the functional prior and posterior is highest at $z\approx 20$ and vanishes at high redshift where the signals are independent of the astrophysics and are defined by the cosmological model thus sharing a common structure.

Panel (b) of \cref{fig:ARad_results} shows the 1D and 2D posterior distributions for the astrophysical parameters $f_*$, $V_c$, $L_X$ and $A_{\mathrm{r}}^{1420}$ when fitting the SARAS3 data with the synchrotron radio background signal model and log-log polynomial foreground model.  From the 1D posterior probability of the parameters, we find that the SARAS3 data disfavours with 68\% confidence a radio background in excess of the CMB by $\gtrsim6\%$ at $1.42$~GHz.
Recent analysis of the HERA power spectrum upper limits using the same set of models disfavoured  a radio background in excess of the CMB of $\gtrsim 1.6$~\% at 1.42~GHz at 68\% confidence. This is a stronger constraint compared to that by SARAS3. However, this claim is model-dependent and assumes that the same astrophysical model that describes CD is valid during the EoR. In reality, the population of sources might evolve and, because HERA is observing at much lower redshifts, it might be probing a different population of astrophysical sources.

Concentrating on the 2D posterior probabilities, we find from the 68\% confidence levels (solid lines  in \cref{fig:ARad_results}) which we approximate by the red dashed lines, that we disfavour different combinations of parameters. Notably, values of $A_\mathrm{r}^{1420} \gtrsim 0.002$ in combination with low values of $f_X \lesssim 43$ are disfavoured. These values correspond to a radio background in excess of the CMB by $\gtrsim 0.2\%$ at $1.4$~GHz and an X-ray luminosity of $ L_\mathrm{X, 0.2 - 95 \textnormal{keV}} \lesssim 1.30\times10^{42} \textnormal{~erg~s}^{-1}$M$_\odot^{-1}$ yr. Similarly, we disfavour  $A_\mathrm{r}^{1420} \gtrsim 9\times10^{-4}$ (an excess of $\gtrsim$0.09\% at 1.42~GHz) in combination with $f_* \gtrsim 5\times10^{-3}$ at approximately 68\% confidence and, separately, high $f_* \gtrsim 0.01$ with low $V_c \lesssim 49$~km/s. The latter constraint can be translated to the limit on the  dark matter halo masses $M \lesssim 3.37\times10^8$~M$_\odot$ at $z=20$.

There are some differences between the constraints for the two considered excess radio background models. For example, the posterior for $L_\mathrm{X}/\mathrm{SFR}$ is flatter when fitting with the high-redshift radio galaxies model in comparison to the synchrotron background model. As highlighted above, the discrepancy  can be attributed to the difference in the evolution of the radio backgrounds with redshift. Since, by construction, the synchrotron radio background decays with time, its impact at higher redshifts is stronger than at lower redshifts. In the case of the radio galaxies model, the radio background builds up with time as stellar populations evolve. Therefore, the constraints are weaker at higher redshifts than at lower.  

\begin{figure}[h!]
    \centering
    \includegraphics{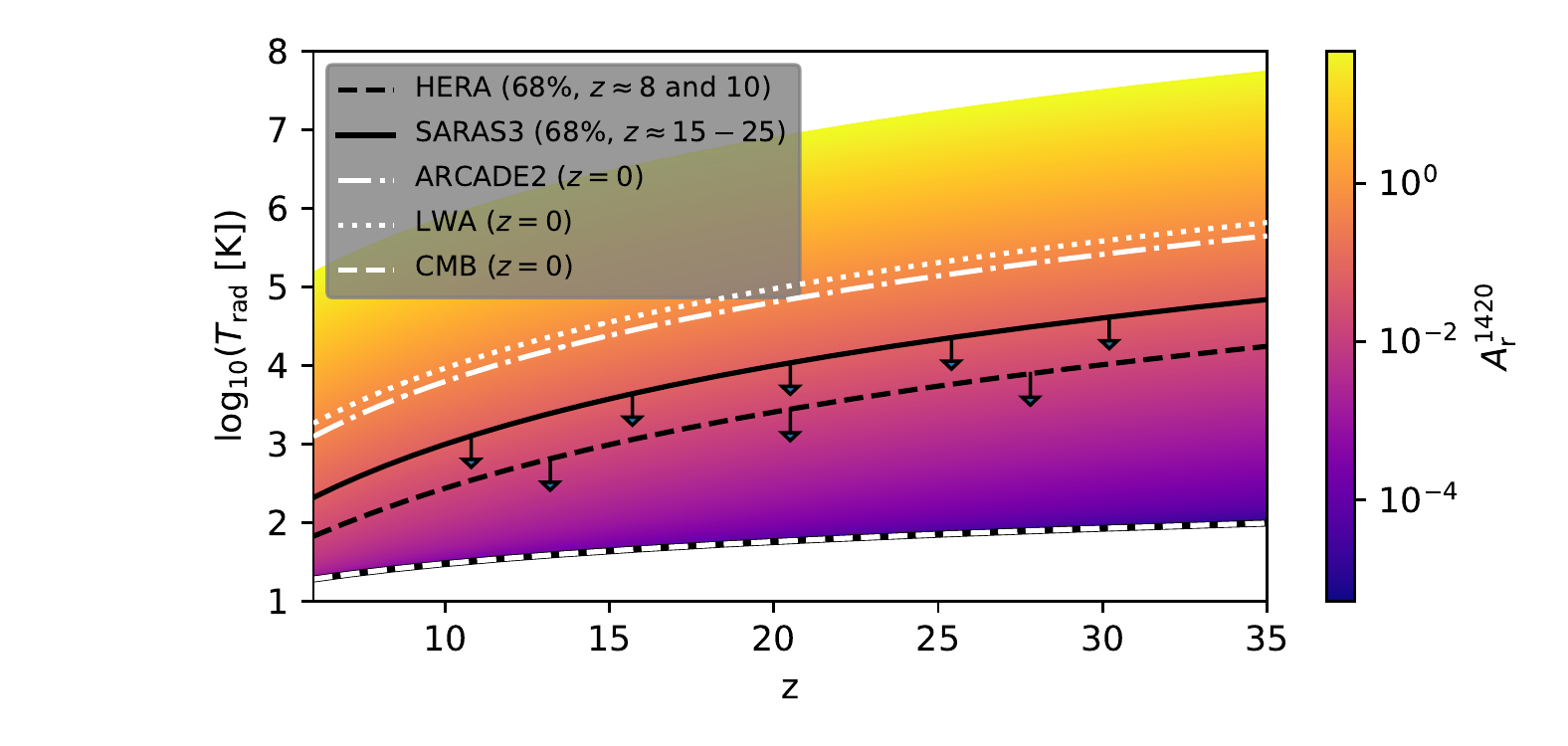}
    \caption{\textbf{The radio background temperature, $T_\mathrm{rad}$, for a set of synchrotron radio background models coloured with respect to their value of $A_{\mathrm{r}}^{1420}$.} We also show the radio backgrounds corresponding to the 68\% confidence limits (black) for SARAS3 extrapolated outside $z\approx 15 - 25$, 68\% confidence limits (black dashed) from HERA extrapolated from $z\approx8$ and $z\approx10$, and the corresponding backgrounds from the ARCADE2 and LWA measurements (white), both extrapolated from $z=0$. We show the CMB temperature for reference.}
    \label{fig:srb_trad}
\end{figure}

Owing to the simple evolution of the radio background temperature, $T_\mathrm{rad}$,   \cref{eq:srb_background},
we can link the SARAS3 68\% constraints at $z\sim 15-25$ to the local measurements of the excess radio background by LWA  \cite{dowell_radio_2018} and ARCADE2 \cite{fixsen_arcade_2011} made at $z=0$ (note, however, the concerns about the Galactic modelling in the ARCADE2 and LWA analysis \cite{Subrahmanyan2013}).  
\Cref{fig:srb_trad} shows the evolution of $T_\mathrm{rad}$   as a function of $z$ and the  value of $A_\mathrm{r}^{1420}$. The solid black line shows the radio background corresponding to the 68\% confidence constraint on $A_{\mathrm{r}}^{1420}$ from the SARAS3 analysis. This limit can be  compared  
to the values observed with ARCADE2  and LWA (shown with white lines). 
We use the reported values of the radio background at 310 MHz, $A_{\mathrm{r}}^{310} = 30.4$ from the LWA measurements \cite{dowell_radio_2018} and $A_{\mathrm{r}}^{310} = 21.1$ from the ARCADE2 measurements \cite{fixsen_arcade_2011}, and scale them appropriately to  78~MHz, assuming a spectral index of $-2.6$, to calculate the corresponding values of $T_\mathrm{rad}$.  The measurements from ARCADE2 and LWA are then extrapolated from $z=0$, while  the SARAS3 constraint is extrapolated outside the band $z\approx 15 - 25$. We show the radio background from the CMB ($A_\mathrm{r}^{1420} = 0$) for reference. Owing to the simplicity of the adopted model, we can connect observations at different redshifts. In reality, as mentioned above, the different experiments may be probing different physical processes because they measure the background at different redshifts and this comparison should be taken as a qualitative one. We find that the SARAS3 limit on the strength of the excess radio background is much lower than the background measured by the ARCADE2 and LWA. Therefore, in this scenario, it is unlikely that the  ARCADE2/LWA measurement is explained by a contribution of high-redshift radio sources.  

Finally, we use the synchrotron radio background models to check the constraining power of the SARAS3 data in the context of the EDGES Low Band detection. We repeat the analysis described in the main text, fitting the data with EDGES-like physical 21-cm signals with a synchrotron radio background global 21-cm signal in combination with the polynomial foreground model. We find that, when selecting for EDGES-like models,  the posterior encapsulates only 57\% of the EDGES-like prior volume. Note that as we are implicitly constraining our prior to produce EDGES-like models, in this piece of analysis and the corresponding analysis in the main text we are assuming a high level of accuracy in the EDGES absorption feature. As a result, the analysis is not intended to `rule out' EDGES-like signals but rather to say that if EDGES is taken at face value as true then 57\% of the corresponding physical parameter space is consistent with SARAS3 (or 43\% is inconsistent).

\subsection{CMB Only Radio Background Models} \label{app:results_sta}

Finally, we consider the standard astrophysical case of the models with the CMB as the only sources of the radio background. Since CMB is a rather weak background, compared to the other models that we considered in this work, there are no deep signals that can be robustly disfavoured with the SARAS3 data. Comparing the 213 mK RMS of the SARAS3 data after foreground subtraction to the relative magnitude of the deepest global signals in this model ~($\sim 165$~mK \cite{Reis_sta_2021}) we expect no strong constraints.  

\begin{figure}[h!]
    \centering
    \includegraphics{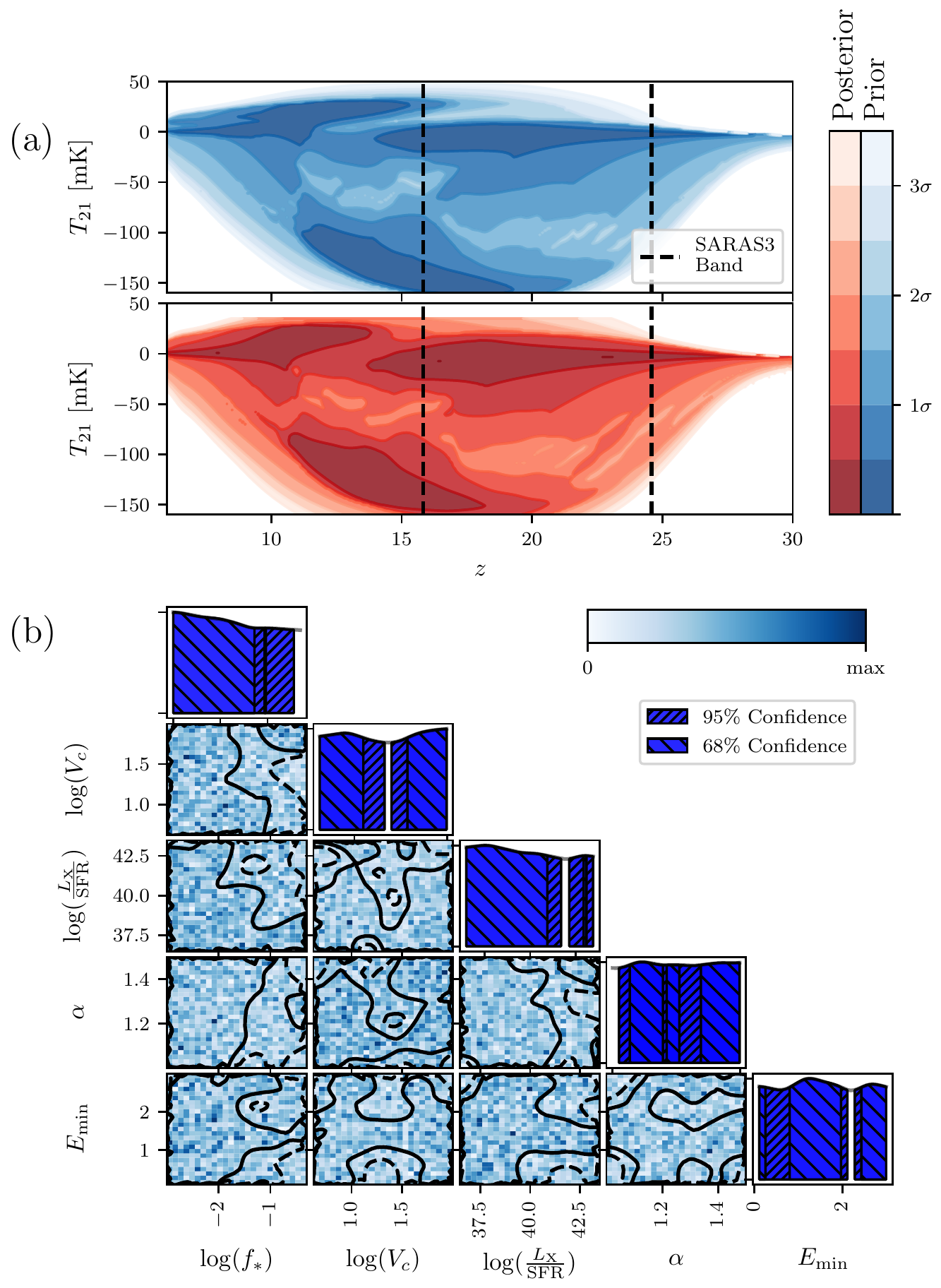}
    \caption{\textbf{SARAS3 constraints on early galaxies assuming a CMB only radio background.}Panel (a) shows the functional prior and posterior found when fitting the SARAS3 data with the CMB only background models, and panel (b) shows the corresponding 1D and 2D posterior distributions for the astrophysical parameters. The distributions are flat and show no constraints, as expected from the magnitude of the SARAS3 residuals after foreground modelling and subtraction.}
    \label{fig:sta_results}
\end{figure}

Indeed, the functional posterior and prior plots in panel (a) of \cref{fig:sta_results} show that there is no significant constraint on the global 21-cm signal for this particular class of models from SARAS3. Further, the corresponding 2D and 1D posteriors, shown in panel (b), exhibit flat distributions and little constraint on the parameter space of the CMB-only models. The models are characterized by a set of astrophysical parameters: the star formation efficiency, $f_*$, the minimum circular velocity of star forming halos, $V_c$, the X-ray efficiency, $f_X$, the slope of the X-ray SED, $\alpha$, the low energy cut-off of the X-ray SED, $E_\mathrm{min}$, the mean free path of ionizing photons, $R_\mathrm{mfp}$, and the CMB optical depth, $\tau$. Again, as SARAS3 is a high-redshift instrument, we marginalized out $\tau$ and for these models we fixed $R_\mathrm{mfp}$ to 30 Mpc.

\subsection{The Functional EDGES-like Posterior and Prior}

\begin{figure}[h!]
    \centering
    \includegraphics{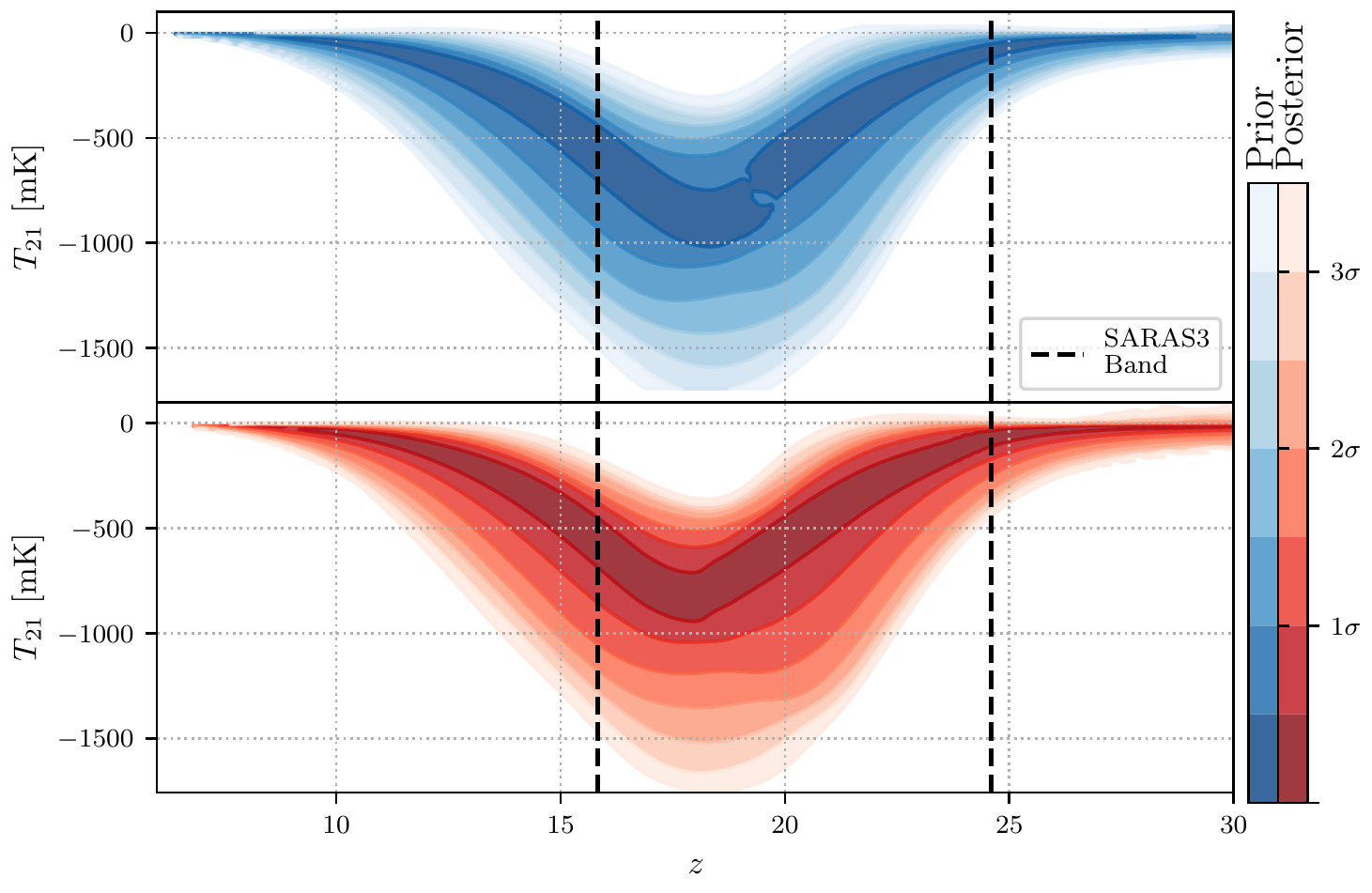}
    \caption{\textbf{The functional EDGES-like prior and the corresponding posterior found when fitting the data using the radio galaxies background models.} The prior is constrained by a conditional equation that selects models with a similar central frequency and depth as the reported EDGES profile \cite{Fialkov2019}.}
    \label{fig:EDGES-like}
\end{figure}

In \cref{fig:EDGES-like}, we show the functional EDGES-like physical posterior (red) and prior (blue) found when fitting the data with the radio galaxies background 21-cm signals. The posterior and prior are used to determine what percentage of the EDGES-like physical signals, those with similar depths and central frequencies to the reported absorption feature, is consistent with the SARAS3 data. This is discussed in detail in the main text and in \textit{Methods}.

\end{document}